## Tutorial: Methods and Software for the Multilevel Social Relations Model

Jeremy Koster[1], George Leckie, Brandy Aven, and Christopher Charlton

# Contents

## Introduction

This tutorial is an introduction to methods and software for fitting the multilevel Social Relations Model (SRM) to cross-sectional, directed network data. The modeling approaches are illustrated via a series of case studies, and the data to replicate these analyses are included as supplemental files (see the link to the Open Science Framework on page 4). The six case studies focus on the SRM for different response types (binary, count, and continuous), as applied either to a single group or multiple groups.

The models in this tutorial are estimated using Stat-JR software developed by the Centre for Multilevel Modeling at the University of Bristol (Charlton et al., 2016). There is a "template" for each version of the SRM, which permits a point-and-click interface for specifying models. Available for downloads at the website of the Centre for Multilevel Modeling, Stat-JR is limited to Microsoft Windows operating systems. The use of Stat-JR requires the additional installation of a C++ compiler, and related hyperlinks and installation instructions for a freely available compiler can be found on the Stat-JR website.[2]

Stat-JR requires that data files be uploaded as Stata datasets (*.dta). In addition to saving files directly from Stata software, there are other options for generating Stata files. For instance, Stat-JR includes a built-in template (*LoadTextFile*) that permits the uploading of text files.[3] Among other options, Stata files can also be created using the *haven* or *foreign* packages in the *R* software environment (the datasets for this tutorial were created using the latter package).

---

[1] Please direct correspondence to jeremy.koster@uc.edu
[2] As of 2019, the hyperlink is: http://www.bristol.ac.uk/cmm/software/statjr/order-statjr/ In our experience, the installation of the compiler is the only noteworthy hurdle to the use of Stat-JR. Registered users are encouraged to request technical support from the CMM staff via the Stat-JR forum if the troubleshooting guide on the website does not resolve problems with the installation.
[3] For more information on this template, see the manual: A Beginner's Guide to Stat-JR's TREE Interface, version 1.0.4.

For all models in this tutorial, the datasets are organized in "long" format. See Table 1 for an example. This represents a vectorization of a sociomatrix in which the actors appear in one column (i_ID) and the partners appear in an adjacent column (j_ID). There is an additional categorical variable that denotes the dyads (ij_ID). This latter variable is symmetric, and each value therefore appears twice in a fully sampled dataset. In other words, the *i* to *j* relation and the *j* to *i* relation both belong to the *ij* dyad. For the SRM models, the Stat-JR templates require this variable to be formatted correctly. Meanwhile, the response variable holds the values for the ties (i.e., corresponding to the values of the cells of the sociomatrix). We also include a vector of 1's, dubbed "Cons" because it is the variable that is supplied to estimate the constant, or the intercept, in the model. Unlike many software packages, the Stat-JR templates do not automatically assume that the model includes a constant, so users must supply this variable.[4]

All of the models in this tutorial are estimated with Markov chain Monte Carlo (MCMC) methods. A full explanation of MCMC methods is beyond the scope of this tutorial (see Browne, 2014 for an introduction in the context of the Stat-JR software). In brief, rather than converging to a point estimate for model parameters, MCMC estimation draws samples from a posterior probability distribution. It is common to draw many samples in order to characterize the distribution of plausible values for the parameter. In practice, this means that for each parameter in the model (including the random effects), thousands of samples are stored, and these samples are then described as the "posterior" distribution. Summary statistics of the posterior, typically the mean and standard deviation of the samples, provide the information that we report for the model parameters. These quantities are analogous to the point estimates and standard errors presented from frequentist analyses.

**Table 1.** Example dataset showing dyadic data in long format, assuming fully sampled data on three nodes (denoted *i*, *j*, and *k*).

| Record | i_ID | j_ID | ij_ID | Response (y) | Cons |
|---|---|---|---|---|---|
| 1 | *i* | *j* | *ij* | $y_{ij}$ | 1 |
| 2 | *i* | *k* | *ik* | $y_{ik}$ | 1 |
| 3 | *j* | *i* | *ij* | $y_{ji}$ | 1 |
| 4 | *j* | *k* | *jk* | $y_{jk}$ | 1 |
| 5 | *k* | *i* | *ik* | $y_{ki}$ | 1 |
| 6 | *k* | *j* | *jk* | $y_{kj}$ | 1 |

[4] For each response type, this tutorial references SRM templates for single groups or multiple groups. In general, the SRM for multiple groups assumes that there are many groups. When there are few groups, researchers might consider using the templates for single groups while using binary indicators as conventional covariates to distinguish group membership and allow for density differences between the groups. The use of fixed effect indicators largely precludes consideration of group-level covariates (such as team performance in the second case study), but this may be an acceptable tradeoff in samples that contain few groups.

The iterated samples in an MCMC chain should exhibit minimal autocorrelation and stationary distributions. In Figure 1, the panel on the left depicts a chain that has these characteristics. It is common, however, for chains to exhibit high autocorrelation and non-stationarity. The panel on the right in Figure 1 depicts a poorly-mixed chain that does not adequately explore the posterior distribution of a parameter. In this latter scenario, a remedy is to lengthen the chain of iterated samples, which allows the algorithm to more thoroughly explore the distribution. For these lengthy chains, which sometimes reach lengths of several million iterations, it is not necessary to store all of the samples. Instead, it is common to "thin" the chains so that every *n*th sample is stored. In general, for the models in this paper, we set the thinning rate at a value that allows us to store 5,000 samples.

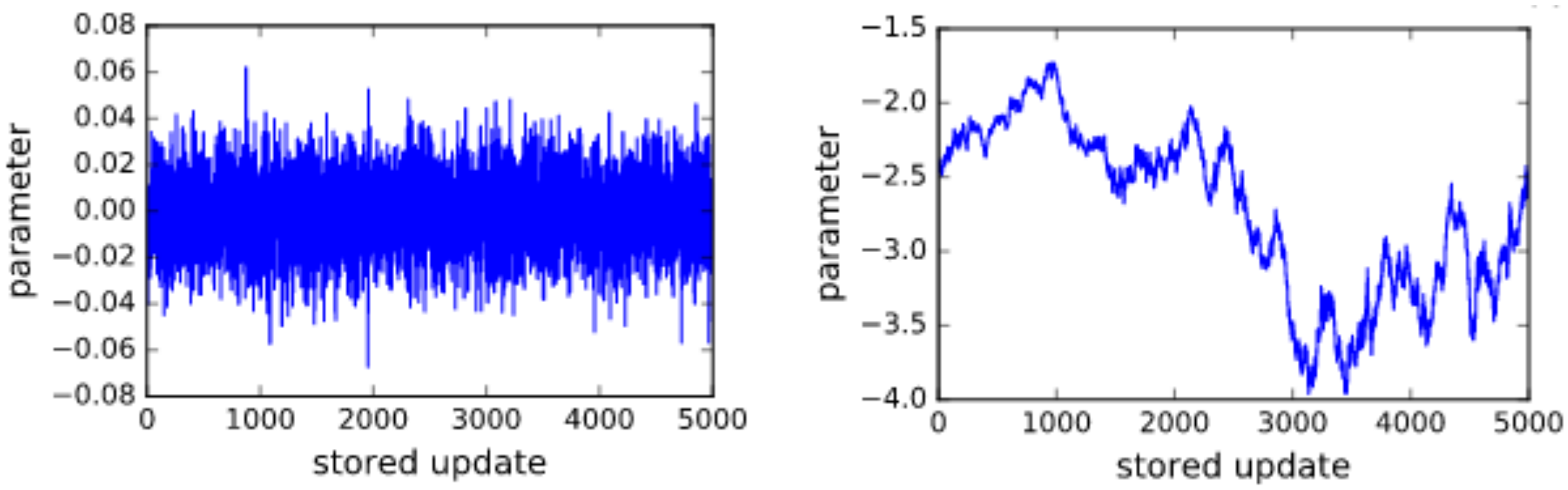

**Figure 1.** Examples of a well-mixed MCMC chain (left) and a poorly-mixed chain that should not be used to summarize a statistical parameter (right).

Another concern with MCMC estimation is that the first samples in the chain have not yet reached the stationary distribution of the parameter. Hence it is common to discard the initial samples from an MCMC chain, which are considered the "burn-in" that allows the algorithm to reach the stationary posterior distribution.

There are no *a priori* guidelines that dictate the necessary lengths for MCMC chains and burn-ins, and researchers should be prepared for some trial-and-error experiences with MCMC chains. In addition to visual inspections of the samples, there are diagnostics available in Stat-JR that allow researchers to assess the autocorrelation and stationarity of their MCMC chains (see Browne et al., 2016:28-31). Note that each parameter is associated with a unique chain of samples, so it is necessary to examine the chains for all parameters. As a final note about MCMC chains, it is possible for posterior distributions to be multi-modal. For models estimated in Stat-JR, we therefore recommend running multiple chains in parallel and then confirming that they all converge to the same posterior distribution. We provide an example in Case Study 2.

This tutorial specifies lengthy MCMC chains, which may require several hours for the models to finish estimating. In general, estimation time increases with the length of chains, the size of the dataset, and the number of parameters to be estimated. For the purpose of familiarizing themselves with the software, users are encouraged to consider specifying substantially shorter chains to confirm that the models are estimating as expected. Note that there are varying opinions on the lengths of chains that are necessary to generate robust results.

Full step-by-step instructions for opening Stat-JR, uploading data and templates, and specifying models are included only for Case Study 1. Researchers who use other templates to specify models should refer to the first case study for guidance. All templates and datasets for the case studies are available via the Open Science Framework (https://osf.io/jkz5t/). File names mentioned in this tutorial match those supplemental files.[5]

The tutorial accompanies a conceptual paper on the multilevel SRM by Koster et al. (*in press*). Because these models are designed primarily for analyzing directed, cross-sectional network ties, it is important to note that the statistical approaches in this tutorial may not be appropriate for modeling other types of dyadic data. For instance, the SRM is not well-suited for the analysis of symmetric dyadic ties, such as friendships on Facebook (Lewis et al., 2008) or organizational co-affiliation (Rider, 2012). Although there have been extensions of the SRM to longitudinal outcomes (Westveld & Hoff, 2011; Nestler et al., 2016), the models in this tutorial could be applied to longitudinal network data only by assuming that the random effects of nodes and dyads do not change over time. Also, the SRM assumes that the same entities (e.g., individuals) appear as both the source of ties (the *i* role) and the recipient of ties (the *j* role). Many dyadic datasets, however, reflect discrete roles in which individuals are observed in only one of those roles, which precludes the estimation of the generalized reciprocity correlation.[6] Finally, the SRM is appropriate for only dyadic outcome variables, not variables in which network ties are used to calculate node-level measures, such as centrality or constraint (Freeman, 1978; Burt, 1992). This generalization applies whether the node-level network measures are the dependent (e.g., Aven, 2015) or the independent variable (Ahuja, 2000; Shipilov and Li, 2008, Aven and Hillmann, 2018).

[5] Users may wish to access datasets and templates across multiple Stat-JR sessions, in which case it is possible to store datasets in their personal data store (C:\Users\<username>\.statjr\datasets) and templates in their personal storage of templates (C:\Users\<username>\.statjr\templates). In that case, the ".dta" and ".py" suffixes would be omitted.

[6] For instance, consider data structures in which multiple judges rate multiple participants in a contest, or multiple auditors monitor the financial reporting of banks (Aven et al. *in press*). In these cases, the dyadic outcome is exclusively unidirectional from one role to the other.

# 1. The Social Relations Model for Binary Responses: Single Group

The first SRM that we demonstrate in this tutorial is appropriate for dyadic data on binary, directed relations from a single group. This version of the SRM uses a probit link function to model the expected probability of a tie between node *i* and node *j*. The model can be written as follows:

$$y_{i,j} = \begin{cases} 1 & y^*_{i,j} \geq 0 \\ 0 & y^*_{i,j} < 0 \end{cases}$$

$$y^*_{i,j} = \mathbf{x}'_{i,j}\boldsymbol{\beta} + a_i + b_j + e_{i,j}$$

$$\begin{pmatrix} a_i \\ b_i \end{pmatrix} \sim N\left\{\begin{pmatrix} 0 \\ 0 \end{pmatrix}, \begin{pmatrix} \sigma^2_a & \\ \sigma_{ab} & \sigma^2_b \end{pmatrix}\right\}, \qquad \rho_{ab} = \frac{\sigma_{ab}}{\sqrt{\sigma^2_a}\sqrt{\sigma^2_b}}$$

$$\begin{pmatrix} e_{i,j} \\ e_{j,i} \end{pmatrix} \sim N\left\{\begin{pmatrix} 0 \\ 0 \end{pmatrix}, \begin{pmatrix} 1 & \\ \sigma_{ee} & 1 \end{pmatrix}\right\}, \qquad \rho_{ee} = \sigma_{ee}$$

One addition to the conventional notation is that we include below the expressions for calculating the variance partition coefficients (VPCs) for actors $(p_a)$, partners $(p_b)$, and dyads $(p_e)$. The Stat-JR templates automate the calculation of VPCs and this information is presented alongside model parameters in the model output.

$$p_a = \frac{\sigma^2_a}{\sigma^2_a + \sigma^2_b + 1}$$

$$p_b = \frac{\sigma^2_b}{\sigma^2_a + \sigma^2_b + 1}$$

$$p_e = \frac{1}{\sigma^2_a + \sigma^2_b + 1}$$

The variance partition coefficient at each level is calculated as the estimated variance at that level divided by the estimated total variance.

## 1.1. Case Study 1: Advice Networks at a High-Tech Firm

The first case study uses data on the advice networks of 33 employees in a high-tech firm, as described by Krackhardt (1999). These employees were organized in a hierarchical structure with supervisors and subordinates (Figure 2). The response variable in this analysis is based on network surveys of the employees in which the individuals were asked to nominate the co-workers to whom they turn for advice. In other words, the variable is whether individual *i* nominates individual *j* as a person who provides them with advice.

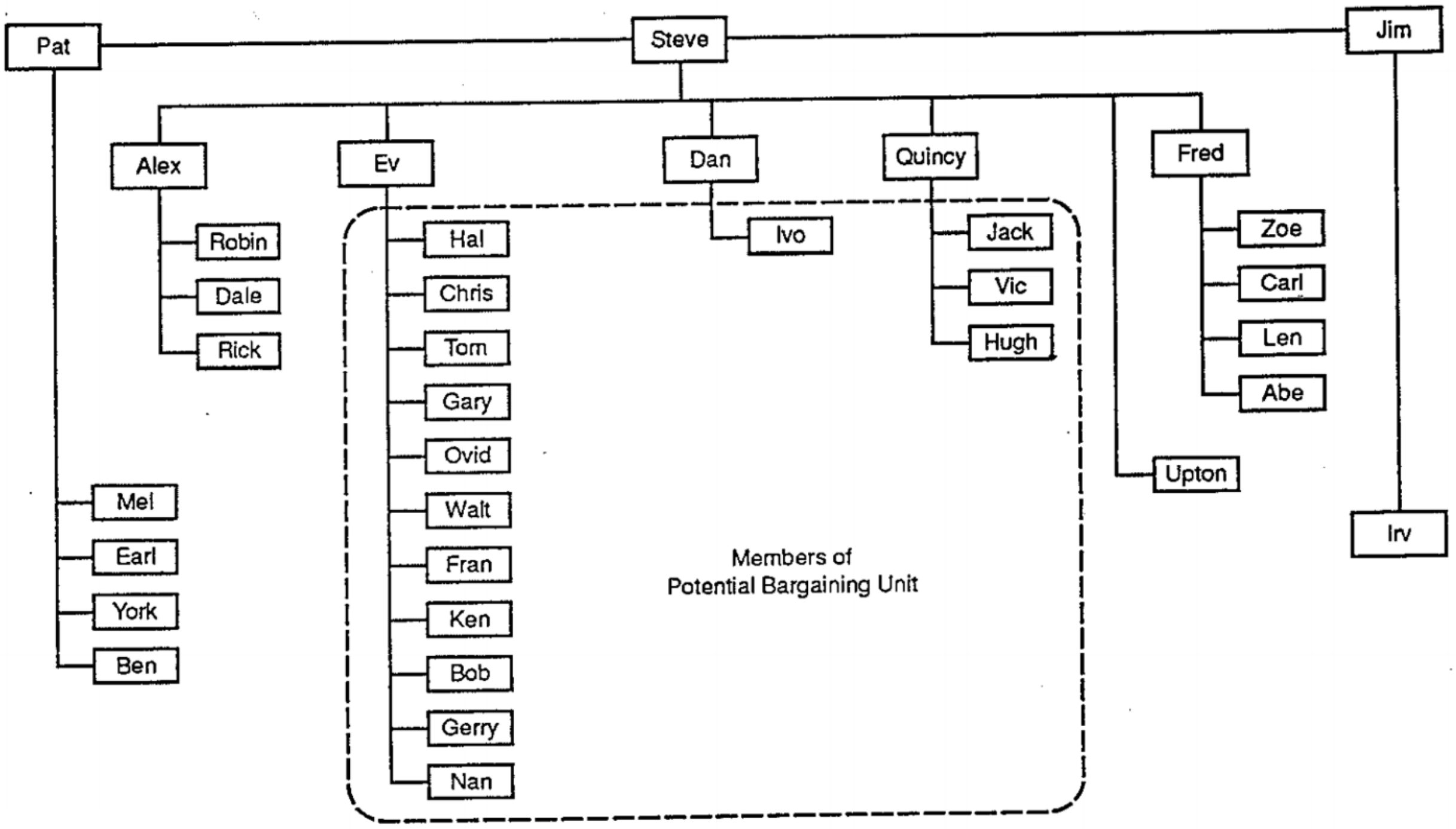

**Figure 2.** The organizational chart at the high-tech firm studied by Krackhardt (1999). Note that Fran, Quincy, and York did not complete the survey and were omitted from the dataset.

In addition to eliciting the advice networks, Krackhardt (1999) also asked the respondents to nominate the co-workers whom they consider friends. On the assumption that co-workers are relatively more likely to ask their friends for advice, we include these friendship ties as a predictor of the advice nominations. This predictor variable is called **ij_friendship**. Other covariates in the analysis are drawn from the organizational chart. We include binary variables to denote when *i* is a direct supervisor of *j* (e.g., Pat is a supervisor of Mel), when *i* reports directly to *j* (e.g., Mel reports to Pat), and when *i* and *j* share the same supervisor (e.g., Mel and Earl both have the same supervisor, Pat). These variables are denoted in the dataset as **ij_j_reports_to_i**, **ij_i_reports_to_j**, and **ij_same_boss**, respectively.

We begin by fitting an "empty" model to the data, which provides insights into the relative variation of the giver, receiver, and dyadic effects. Subsequently, we include the above covariates to examine the extent to which they predict the advice ties. The dataset is named **1_Binary_one_group_high_tech.dta**, and the file name of the Stat-JR template to analyze the data is **1_Binary_one_group.py**.

For the empty model using the Stat-JR template for binary, single groups, follow these steps.

1. Open Stat-JR – TREE application, which will appear as a tab in the default web browser.
2. Click on the **Dataset** drop-down menu, click **Upload**, then navigate to the location of the supplemental file, **1_Binary_one_group_high_tech.dta**.
3. Again click on the **Dataset** drop-down menu, click **Choose**, and select the **1_Binary_one_group_high_tech.dta** from among the options in the drop-down list. Click the **Use** button.
4. Click on the **Template** drop-down menu, click **Upload**, then navigate to the location of the template, **1_Binary_one_group.py**.
5. Again click on the **Template** menu, click **Choose**, then select the template, **1_Binary_one_group.py**. The screen will change to include new drop-down options.
6. For **Actor ID**, select the variable, **i_ID**. Click the **Next** button.
7. For **Partner ID**, select the variable, **j_ID**. Click the **Next** button.
8. For **Dyad ID**, select the symmetric index variable, **ij_ID**.
9. For **Response**, select the variable, **y**. Click the **Next** button.
10. For **Covariates**, select the variable, **cons**. Click the **Next** button.
11. For the **Prior distribution for actor partner covariance matrix**, we recommend selecting the **Wishart** prior. Then click **Next**.
12. For the **Prior guess for actor partner covariance matrix**, supply the following values in the box, with commas: **0.5,0,0.5**
13. For **Degrees of freedom for Wishart distribution**, supply the value **2.** Then click **Next**.
14. The **Number of Chains** is potentially variable according to researchers' interests. To replicate the case study that we present, select **3**.
15. For the **Random Seed**, select **1** (again noting that alternative values could be supplied).
16. For the **Length of burnin**, select **50000**. (See note above that burn-in lengths can be variable depending on diagnostics conducted by the researcher. Our preliminary modeling suggested that 50,000 iterations were more than sufficient for the chains to reach their stationary distributions.)
17. For **Number of Iterations**, select **100000**. (As with the burn-in, we found that this was sufficient for stationary parameter chains in all models from this case study, including the subsequent model with covariates.)
18. For **Thinning**, select **20**. (Combined with the length of the chain, 100 thousand iterations, a thinning of 20 will result in 5,000 stored iterations for each chain.)
19. For **Use default algorithm settings**, click **Yes**. Then click **Next**. (The other options are recommended for use primarily by advanced users.)
20. For **Generate prediction dataset**, click **No**. (There is no reason that a prediction dataset could not be generated – it is simply that is not emphasized in this tutorial.)
21. For **Use default starting values**, click **Yes**. Then click **Next**.
22. For **Impute at iterations**, select 0
23. For **Name of output results**, one must choose a name for the file that will contain the posterior samples. We choose to call this file **Model 1A**. Then click **Next**.
24. The software will indicate that it is initializing. Clicking **Run** will start the MCMC chain.

Following these steps, the settings screen will appear as follows:

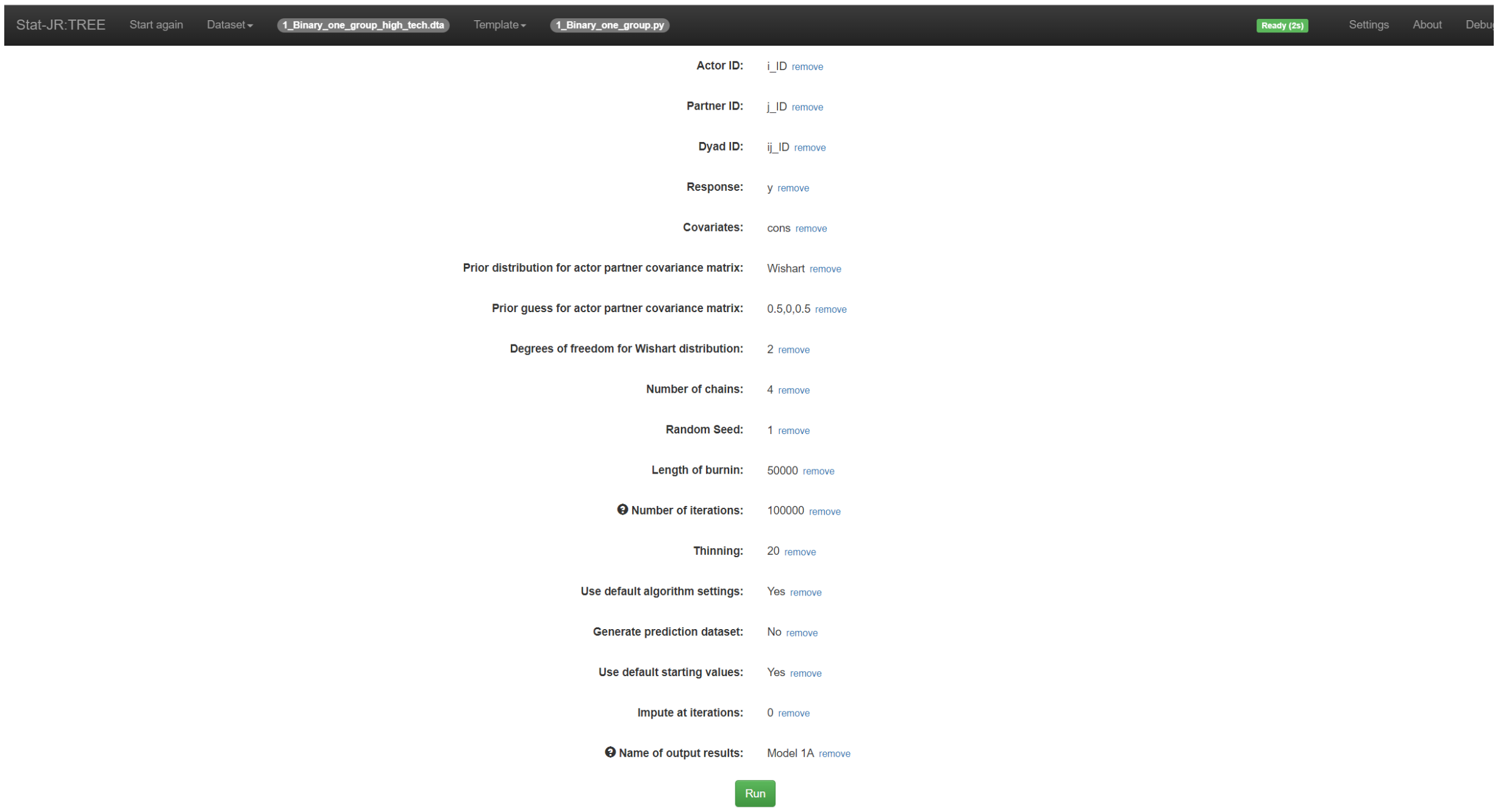

Once the model has finished estimating a number of options for reviewing the model will appear in the drop-down box on the lower left of the screen (the default option is **equation.tex**).[7] Clicking on **ModelParameters** will bring up a table with the following values:

| parameter | mean | sd | ESS | variable |
|---|---|---|---|---|
| **sigma2b1** | 1.03177255313 | 0.437324199139 | 3339 | |
| **pb1** | 0.420688990593 | 0.0884875171843 | 4146 | |
| **rhoe1e1** | 0.732137286779 | 0.11555744034 | 1223 | |
| **sigma2a1** | 0.3346144085 | 0.12847533499 | 10188 | |
| **pe1** | 0.437860343629 | 0.0779209912165 | 3888 | |
| **sigma2r** | 2.36638696163 | 0.485950671065 | 3405 | |
| **beta_0** | -1.83850984968 | 0.266383916977 | 3114 | cons |
| **pa1** | 0.141450665778 | 0.0444955953462 | 12123 | |
| **deviance** | 2536.05512754 | 234.718752166 | 874 | |
| **rhoa1b1** | 0.342345275788 | 0.20929377575 | 9943 | |

**Table 2.** Model parameters, as generated by Stat-JR template for Model 1A.

This table contains the values that will be of primary interest to most applied researchers. As the software is generic and is unaware of any specific meaning for the parameter names, the

[7] As a suggestion, it may be preferable for users learning the software to specify much shorter chains of only a few thousand iterations to confirm that the models are being estimated as expected. With shorter chains, the models typically finish estimating within several minutes.

table of parameters is organized unusually; for example, the table does not list all of the variances together, then the reciprocity correlations, then the variance partition coefficients, etc. We recognize that new users might be confused by the structure of the parameters table, so in Appendix A of this tutorial we provide a key to the parameter names for all of the SRM templates. In general, the *sigma* parameters are the variances of actors (*sigma2a1*) and partners (*sigma2b1*). Recall from the notation above that the dyadic variance is constrained to 1, so it is not reported in the table. The parameter, *sigma2r*, represents the sum of the three variances (actors, partners, and dyads), which is approximately 2.37, and this value serves as the denominator for the calculation of the variance partition coefficients: *pa1*, *pb1*, and *pe*1, respectively. The reciprocity correlations are *rhoa1b1* for generalized reciprocity and *rho1e1* for dyadic reciprocity.

The empty model includes a single fixed effect covariate, the intercept (*beta_0*). Because this model uses a probit parameterization, the coefficients can be converted to the probability scale via the cumulative distribution function of the standard normal. We provide additional details in the second case study. As a reminder, the table provides the mean and standard deviation of the posterior samples. These are generally analogous to the point estimates and standard errors from maximum likelihood models; hence the regression coefficients may be regarded as "statistically significant" different from zero when the absolute value of its *z*-ratio (the quotient of the mean divided by the standard deviation) exceeds a conventional threshold, such as the threshold of 1.96 to achieve a *p*-value of 0.05.

The table also includes the effective sample size (ESS) for each of the parameters. This value provides diagnostic information about the mixing of the chains. If this number is relatively close to the number of stored samples, then it indicates a well-mixing chain. We had stored 5,000 samples from four separate chains, so there are a total of 20,000 stored samples. The ESS for the parameters exceeds 1,000, which generally indicates acceptable mixing.

Other diagnostics can be accessed via the drop-down box. For example, to see the MCMC chain for the dyadic reciprocity correlation, select *rhoe1e1.svg*, which reveals several diagnostic plots for this parameter (Figure 3). All chains exhibit good mixing, and the smoothed histograms of the samples show that all four chains converged on essentially the same distribution. The plots in the middle row depict the amount of autocorrelation in the samples, which is minimal in this case. Overall, these diagnostic plots are reassuring that the posterior samples adequately reflect the distribution.

The interpretation of the empty model shows that there is more variance among employees who provide advice ($\sigma_b^2 = 1.03$) than among employees' reported advice-seeking ties ($\sigma_a^2 = 0.33$). In other words, employees generally nominate a similar number of alters who provide advice, but there is more variability in how many nominations a particular employee receives from co-workers. There is also considerable evidence for dyadic effects, as the VPC shows that about 43% of the variance is evident among dyads.[8] The generalized reciprocity correlation is positive ($\rho_{ab} = 0.34$), but estimated somewhat imprecisely, as seen in the posterior standard deviation for this parameter (0.21). In the parlance of social network analysis, this implies a weak correlation between in-degree and out-degree centrality (Freeman, 1978). By contrast, there is high dyadic reciprocity ($\rho_{ee} = 0.73$). This implies that if individual *i* provides advice to individual *j*, there is a high probability that *j* also provides advice to *i*.

[8] In these models, the dyad-level variance is not reported because it is constrained to a value of 1 to identify the model. As seen in the notation above, however, this value can be used as the numerator and divided by the sum of the variances to calculate the VPC for the dyadic variance.

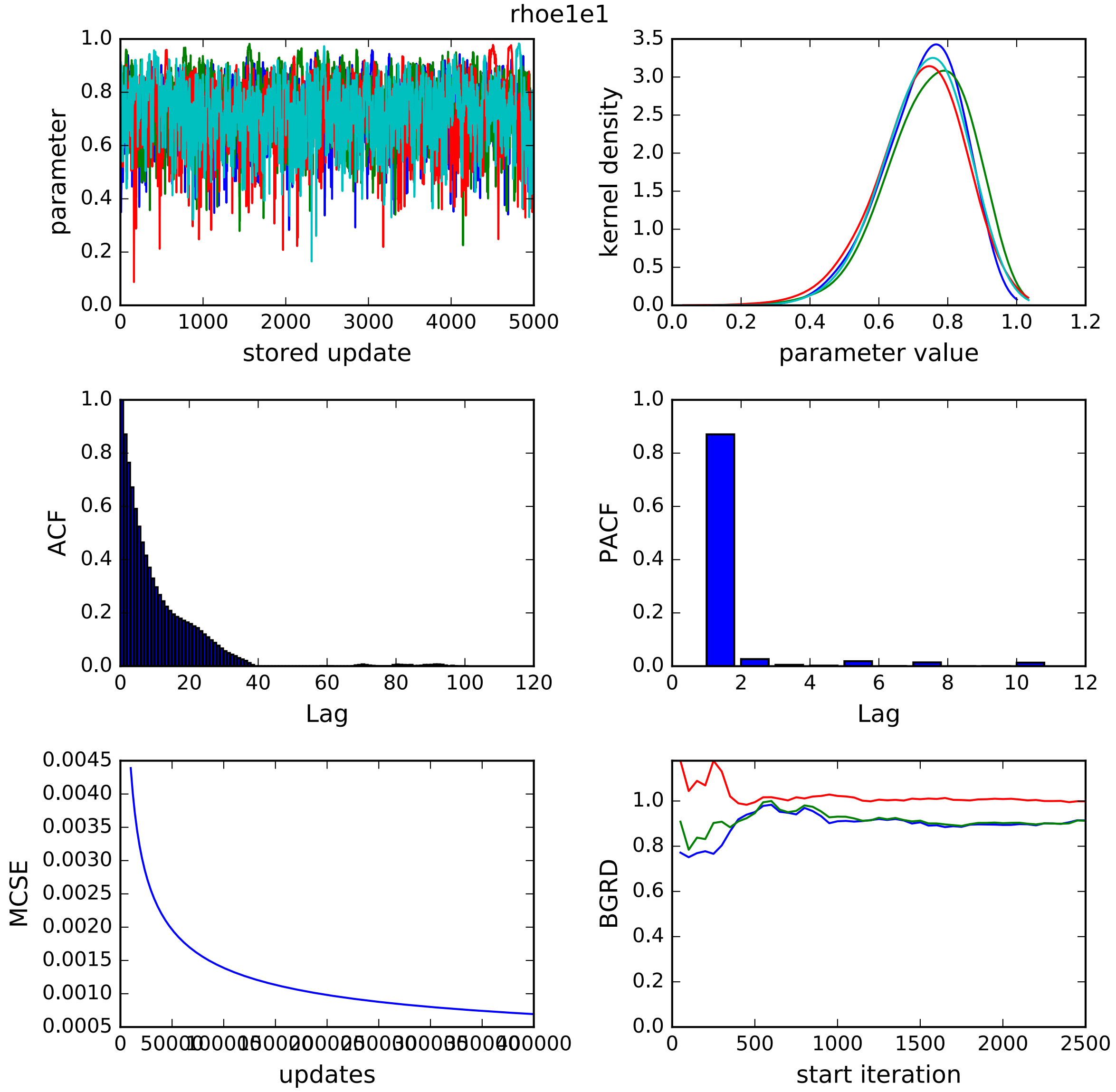

**Figure 3.** Diagnostic plots for the dyadic reciprocity parameter in Model 1A. Similar diagnostic plots are available for other parameters in the model.

Adding further covariates to the model is straightforward. To specify Model 1B, follow these steps.

1. Click on **remove** next to the **cons** in the **Covariates** box. This removes all covariates. Now add the following predictor variables: **cons, ij_same_boss, ij_friendship, ij_i_reports_to_j, ij_j_reports_to_i.**
2. Then repeat all of the steps above from fitting Model 1A.
3. For **Name of output results**, choose **Model 1B.**

This model, which exhibits generally good mixing, yields the parameters in Table 3 (which is reproduced from the Model Parameters table, as above). The positive effect for **ij_friendship** indicates that employees are more likely to seek advice from co-workers whom they consider friends. The positive effect for **ij_i_reports_to_j** indicates that employees show tendencies to seek advice from their supervisors. Conversely, the positive effect for **ij_j_reports_to_i** shows that supervisors are also more likely to seek advice from the individuals whom they oversee rather than others in the organization. The variable for **ij_same_boss** provides evidence that employees are relatively more likely to seek advice from co-workers who share the same functional role for the company.

After adding variables to models, it is worthwhile to consider how the estimated variances and covariances have changed. Notably, the variance for individual *i* has increased ($\sigma_a^2 = 0.72$), as has the variance for individual $j$ ($\sigma_b^2 = 1.37$). This implies that although in general the organizational chart and friendship ties help to explain variation in advice relations, there are some individuals who stand out in terms of seeking and providing advice in ways that are not captured by our covariates. Furthermore, there is moderately stronger statistical support for the generalized reciprocity correlation ($\rho_{ab} = 0.42$). By contrast, the dyadic reciprocity correlation remains strongly positive, though modestly weaker ($\rho_{ee} = 0.64$). The opportunity to evaluate changes to these correlations with the inclusion of predictor variables is a noteworthy advantage of the multilevel Social Relations Model.

| **Stat-JR Parameter** | **Parameter** | **Posterior Mean** | **Posterior SD** | **ESS** |
|---|---|---|---|---|
| beta_0 | Intercept | -2.653 | 0.381 | 1784 |
| beta_1 | ij_same_boss | 0.992 | 0.259 | 7534 |
| beta_2 | ij_friendship | 1.215 | 0.215 | 10975 |
| beta_3 | ij_i_reports_to_j | 2.673 | 0.452 | 9480 |
| beta_4 | ij_j_reports_to_i | 1.050 | 0.446 | 11496 |
| sigma2a1 | Actor *i* variance | 0.716 | 0.301 | 5048 |
| sigma2b1 | Partner *j* variance | 1.370 | 0.627 | 2139 |
| rhoa1b1 | Generalized reciprocity | 0.417 | 0.202 | 8353 |
| rhoe1e1 | Dyadic reciprocity | 0.641 | 0.155 | 1183 |
| pa1 | VPC Actor *i* | 0.230 | 0.069 | 7369 |
| pb1 | VPC Partner *j* | 0.428 | 0.095 | 3367 |
| pe1 | VPC Dyad | 0.342 | 0.076 | 2354 |
| sigma2r | Sum variance | 3.086 | 0.775 | 2065 |

**Table 3.** Model parameters from Model 1B. The column "Stat-JR Parameter," corresponds to the names of parameters in the output file, as seen in the example above. The parameters have been reorganized into an alternative order for the sake of presentation in this table.

## 2. The Social Relations Model for Binary Responses: Multiple Groups

The second SRM that we demonstrate in this tutorial is appropriate for dyadic data on binary, directed relations from multiple groups. As in the first case study, this version of the SRM uses a probit link function to model the expected probability of a dichotomous tie between node $i$ and node $j$. Distinguished from the preceding model by the inclusion of group-level random effects ($m_k$), the model can be notated as follows:

$$y_{ik,jk} = \begin{cases} 1 & y^*_{ik,jk} \geq 0 \\ 0 & y^*_{ik,jk} < 0 \end{cases}$$

$$y^*_{ik,jk} = \mathbf{x}'_{ik,jk}\boldsymbol{\beta} + m_k + a_{ik} + b_{jk} + e_{ik,jk}$$

$$m_k \sim N(0, \sigma^2_m)$$

$$\begin{pmatrix} a_{ik} \\ b_{ik} \end{pmatrix} \sim N\left\{\begin{pmatrix} 0 \\ 0 \end{pmatrix}, \begin{pmatrix} \sigma^2_a & \\ \sigma_{ab} & \sigma^2_b \end{pmatrix}\right\}, \qquad \rho_{ab} = \frac{\sigma_{ab}}{\sqrt{\sigma^2_a}\sqrt{\sigma^2_b}}$$

$$\begin{pmatrix} e_{ik,jk} \\ e_{jk,ik} \end{pmatrix} \sim N\left\{\begin{pmatrix} 0 \\ 0 \end{pmatrix}, \begin{pmatrix} 1 & \\ \sigma_{ee} & 1 \end{pmatrix}\right\}, \qquad \rho_{ee} = \sigma_{ee}$$

$$p_m = \frac{\sigma^2_m}{\sigma^2_m + \sigma^2_a + \sigma^2_b + 1}$$

$$p_a = \frac{\sigma^2_a}{\sigma^2_m + \sigma^2_a + \sigma^2_b + 1}$$

$$p_b = \frac{\sigma^2_b}{\sigma^2_m + \sigma^2_a + \sigma^2_b + 1}$$

$$p_e = \frac{1}{\sigma^2_m + \sigma^2_a + \sigma^2_b + 1}$$

Again, the notation includes the calculations of the VPCs.

### 2.1. Case Study 1: Twitter following among NBA teammates

The second case study draws on data about "following" relationships on Twitter among NBA players (Koster and Aven, 2018). At the time of data collection in May 2015, there were 330 players who maintained Twitter accounts, and these players were distributed among 30 teams for a total sample size of 3,356 possible following relationships. The response variable is whether player $i$ follows the Twitter account of teammate $j$. The dataset, called **2_Binary_multi_group_NBA_twitter.dta**, includes several covariates and interaction terms that potentially predict following ties. These predictors include player-level variables such as the number of All-Star nominations they received, dyadic variables such as the number of seasons the players have been teammates, and team-level variables such as the recent performance of the team in the postseason playoffs (for full details, see Koster and Aven, 2018). The nomenclature for distinguishing actors, partners, and dyads is consistent with the first case study (**i_ID**, **j_ID**, **ij_ID**). The variable, **k_ID**, is used to distinguish different NBA teams. The template for fitting these data is called **2_Binary_multi_group.py.**

The process for fitting an empty model replicates the steps taken in the first case study, and the specified model appears as follows in the Stat-JR interface:

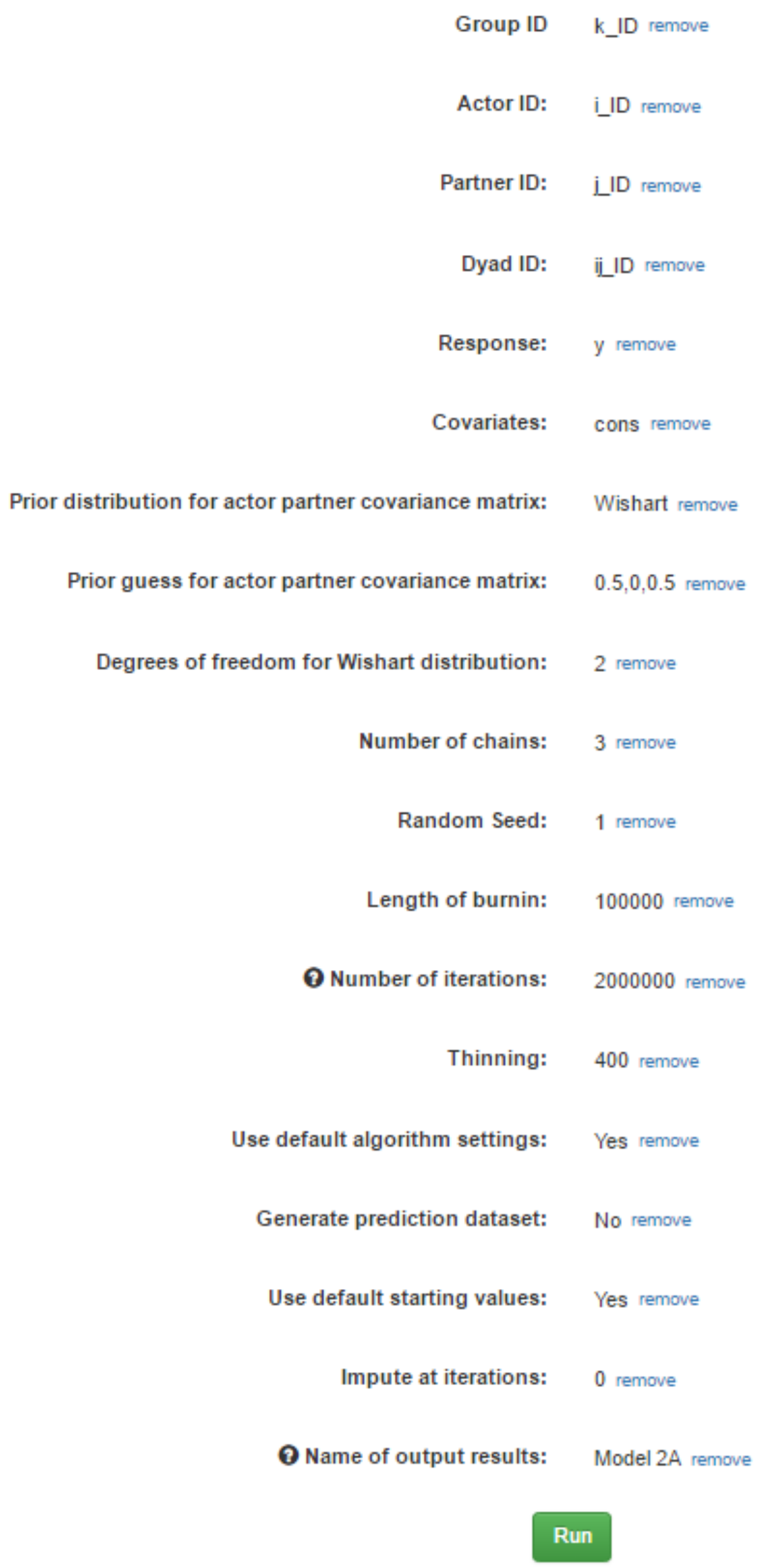

Models were estimated using a burn-in of 100,000 iterations and 2,000,000 monitoring iterations.

In addition to an empty model (Model 2A), we estimate a model that includes covariates (Model 2B), which are added by following steps described above (Table 4). Many of these variables have been *z*-score or proportionally standardized, which tends to facilitate better mixing of the MCMC chains than the use of unstandardized variables (particularly those with large ranges).

| **Parameter** | **Description** | **Model 2A** | **Model 2B** |
|---|---|---|---|
| beta_0 | Intercept | 0.130 (0.128) | -0.841 (0.168) |
| beta_1 | League tenure, player *i* (logged and *z*-score) | | -0.275 (0.091) |
| beta_2 | League tenure, player *j* (logged and *z*-score) | | -0.170 (0.064) |
| beta_3 | League tenure *i* * league tenure *i* | | 0.121 (0.041) |
| beta_4 | Salary, player *i* (logged and *z*-score) | | 0.167 (0.086) |
| beta_5 | Salary, player *j* (logged and *z*-score) | | 0.328 (0.061) |
| beta_6 | Salary *i* * Salary *j* | | 0.058 (0.038) |
| beta_7 | Time using Twitter, player *i* (*z*-score) | | 0.183 (0.076) |
| beta_8 | Time using Twitter, player *j* (*z*-score) | | 0.234 (0.053) |
| beta_9 | Time using Twitter *i* * Time using Twitter *j* | | 0.042 (0.038) |
| beta_10 | Height, player *i* (*z*-score) | | 0.096 (0.071) |
| beta_11 | Height, player *j* (*z*-score) | | 0.040 (0.049) |
| beta_12 | Height *i* * Height *j* | | -0.02 (0.036) |
| beta_13 | Years as teammates (proportionally standardized) | | 8.671 (0.870) |
| beta_14 | Same college | | 0.775 (0.308) |
| beta_15 | All-star games, player *i* (proportionally standardized) | | -1.848 (0.885) |
| beta_16 | All-star games, player *j* (proportionally standardized) | | 0.291 (0.601) |
| beta_17 | Playoff performance, team *k* (*z*-score) | | 0.192 (0.144) |
| beta_18 | All-star games *i* * All-star games *j* | | -3.718 (6.46) |
| beta_19 | All-star games *i* * team playoff performance | | 1.655 (0.811) |
| beta_20 | All-star games *j* * team playoff performance | | 0.260 (0.575) |
| beta_21 | All-star games *i* * All-star games *j* * team playoff performance | | 4.205 (7.853) |
| sigma2m | Variance of teams | 0.155 (0.151) | 0.208 (0.172) |
| sigma2a1 | Variance of player *i* | 1.169 (0.15) | 1.290 (0.173) |
| sigma2b1 | Variance of teammate *j* | 0.706 (0.099) | 0.496 (0.08) |
| rhoa1b1 | Generalized reciprocity correlation | 0.751 (0.038) | 0.793 (0.038) |
| rhoe1e1 | Dyadic reciprocity correlation | 0.902 (0.027) | 0.881 (0.032) |
| pm | VPC, team *k* | 0.050 (0.045) | 0.068 (0.052) |
| pa1 | VPC, player *i* | 0.385 (0.032) | 0.430 (0.037) |
| pb1 | VPC, teammate *j* | 0.233 (0.023) | 0.166 (0.021) |
| pe1 | VPC, dyad *ij* | 0.332 (0.027) | 0.336 (0.03) |
| sigma2r | Sum variance | 3.029 (0.252) | 2.996 (0.27) |

**Table 4.** Model parameters from Model 2A and 2B. Asterisks denote interaction terms.

As seen in the empty model, Model 2A, the main contrast with the SRM for single groups is the addition of random effects for the groups. In this case, the variance of effects for NBA teams is modest, ($\sigma_m^2 = 0.16$), accounting for only about 5% of the total variance. In the terminology of social network analysis, this implies that the density of ties varies relatively little from team to team. The other parameters can be interpreted as in the single-group model. There is substantial variation in the players' tendencies to follow teammates ($\sigma_a^2 = 1.17$). In addition, some players are followed by more teammates than others ($\sigma_b^2 = 0.71$). The generalized reciprocity is high ($\rho_{ab} = 0.75$). Similarly, the dyadic reciprocity correlation is very high ($\rho_{ab} = 0.90$), implying that Twitter following ties are typically reciprocated. These correlations remain largely consistent with the inclusion of covariates in Model 2B.

## 2.2. Predicted probabilities and customized predictions

For models fitted with a probit link function, the coefficients are on the scale of the cumulative normal distribution.[9] In terms of calculating probabilities for multilevel models, including the probit models in our first two case studies, different methods have been proposed (Steele, 2009). One of these methods, which generates "cluster-specific" predictions, ignores the variability related to the random effects by assuming average random effects in all cases – recall that random effects in these are assumed to be distributed around a mean of zero. This is the method used at present to generate predicted probabilities. An alternative is to simulate values from the distribution of the random effects and then use those simulated values in calculations of model predictions (Steele, 2009), but we do not explore this alternative approach further here.

Calculating the cluster-specific predicted probability from Model 2A is relatively straightforward, as one compares the estimated intercept (0.13) to the standard normal distribution, yielding a predicted probability of 0.55 for Twitter following ties among teammates. This is very close to the proportion of ties in the empirical dataset (0.54), with slight deviations to be expected by virtue of ignoring the random effects.

When calculating the predicted cluster-specific probabilities from a model with multiple fixed effect covariates, one must supply values that can be multiplied by the coefficients. For example, imagine that we wish to calculate the predicted probabilities of Twitter ties for NBA teammates who either attended the same college or not (**Same college** is a binary indicator variable). For all other predictors in the model, we can supply the average values from the empirical dataset. In lieu of equations, a table potentially makes these calculations clearer (Table 5). The calculations in this case are fairly straightforward in part because most of the variables have been *z*-score standardized. We can therefore supply values of 0 for these predictors, which corresponds to the average value of that variable. The only predictor variable that, by definition, requires a non-zero value is **ij_years_mates_prop**. This is a proportionally standardized variable (relative to the maximum of 13 seasons) that denotes how many years player *i* and player *j* have been teammates. Since the dataset includes only players who have been teammates during at least one season, a non-zero value is not permitted. Accordingly, we supply a value of 0.08, which corresponds to a value of one season that the players have been teammates (1/13 = 0.08).

[9] This distribution will be familiar to some researchers because it is the distribution used to calculate *p*-values from a *z*-score.

| Variable | Posterior mean | Prediction Value 1 | Product 1 | Prediction Value 2 | Product 2 |
|---|---|---|---|---|---|
| cons | -0.84 | 1 | -0.84 | 1 | -0.84 |
| i_tenure_log_z | -0.27 | 0 | 0 | 0 | 0 |
| j_tenure_log_z | -0.17 | 0 | 0 | 0 | 0 |
| ij_tenure_interaction | 0.12 | 0 | 0 | 0 | 0 |
| i_salary_log_z | 0.17 | 0 | 0 | 0 | 0 |
| j_salary_log_z | 0.33 | 0 | 0 | 0 | 0 |
| ij_salary_interaction | 0.06 | 0 | 0 | 0 | 0 |
| i_time_z | 0.18 | 0 | 0 | 0 | 0 |
| j_time_z | 0.23 | 0 | 0 | 0 | 0 |
| ij_time_interaction | 0.04 | 0 | 0 | 0 | 0 |
| i_height_z | 0.10 | 0 | 0 | 0 | 0 |
| j_height_z | 0.04 | 0 | 0 | 0 | 0 |
| ij_height_interaction | -0.02 | 0 | 0 | 0 | 0 |
| ij_years_mates_prop | 8.67 | 0.08 | 0.67 | 0.08 | 0.67 |
| ij_college | 0.77 | 0 | 0 | 1 | 0.77 |
| i_stars_prop | -1.85 | 0 | 0 | 0 | 0 |
| j_stars_prop | 0.29 | 0 | 0 | 0 | 0 |
| k_wins_sqrt_z | 0.19 | 0 | 0 | 0 | 0 |
| ij_stars_interaction | -3.72 | 0 | 0 | 0 | 0 |
| ik_stars_wins_interaction | 1.66 | 0 | 0 | 0 | 0 |
| jk_stars_wins_interaction | 0.26 | 0 | 0 | 0 | 0 |
| ijk_stars_wins_interaction | 4.21 | 0 | 0 | 0 | 0 |
| | | **Sum 1** | **-0.17** | **Sum 2** | **0.60** |
| | | **Probability 1** | **0.43** | **Probability 2** | **0.73** |

**Table 5.** Demonstration of calculating predicted probabilities for Model 2B, supplying means or reference values for nearly all coefficients except for **ij_college**, which is first supplied a value of 0 and subsequently a value of 1 to determine the predicted effect of having attended the same college. The columns, Product 1 and Product 2, represent the product of the posterior means for the variables and the values that are supplied in Prediction Value 1 and Prediction Value 2, respectively.

In the first calculation, the sum of the products of the supplied values and the posterior means is −0.17, which corresponds to a predicted probability of 0.43 for a following tie. In the second calculation, the products sum to 0.60, which corresponds to a predicted probability of 0.73 in the standard normal distribution. Thus, the interpretation of this comparison is that, holding constant all other parameters at their respective values, the model predicts an increase in the probability of a following tie from the baseline of 43% to 73% among teammates who formerly attended the same college. Similar calculations can generate predicted probabilities for other values that are supplied by the researcher.

The above calculations, however, consider only the mean values from the posterior distributions of the parameters. There are 20,000 stored samples for each parameter, and each is considered an independent realization from the posterior.[10] For each iteration in the chain, that is, the parameter values fluctuate around the posterior mean (as evidenced earlier by the diagnostic plots of the MCMC chains). Researchers who wish to contextualize the uncertainty in their model estimates can recalculate the predicted probabilities for each posterior sample. These calculations then lend themselves to summarizing the mean and credibility intervals of the model predictions.

For example, Figure 4 depicts the predictions from Model 2B while allowing for variation in the All-Star status of player *i*. Notably, this variable is interacted in the model with the All-Star status of player *j* and a measure of team performance. To generate this plot, first a vector of values of length *n* was created that encompasses the range of team performance. Then we set up prediction sets that assumed player *i* was either a four-time All-Star or a player who had never participated in an All-Star game. We assumed that player *i* had four years of tenure, and we held other variables constant at their mean or reference levels. Finally, for each of the two customized datasets, we multiplied the *n* set of parameter values by their corresponding coefficients in each of the 20,000 stored samples from the posterior distribution. This generates 20,000 predicted probabilities along the range of the vector of length *n*, and then it is fairly straightforward to calculate the mean and confidence intervals around the predictions.[11]

Figure 4 was generated in the R software environment, and the coding script to replicate this figure is included in Appendix B. The script requires two input files: (1) the empirical data in the file, **2_Binary_multi_group_NBA_twitter.dta** and (2) the file containing the posterior samples, **Model 2B.dta**, which is the name that had been assigned to the output results when setting up the model in Stat-JR. When the model has finished estimating, clicking the green **Download** button in Stat-JR generates a zipped file that contains the output file with the posterior samples (among other files). This file can then be read into other statistical programs to facilitate other calculations from the posterior samples. The script in Appendix B uses the R software environment, but similar calculations could be accomplished with most general-purpose statistical software.

---

[10] That is, there were 4 chains with 5,000 stored samples in each chain, yielding a total of 20,000 stored samples.

[11] Note that even though mean values were supplied for other coefficients in the model, those coefficients vary from sample to sample in the posterior. Hence the distribution of predicted values reflects the uncertainty in all parameters, not just the parameters of interest (the coefficients for players' salaries in this example). Confidence intervals may overlap even when a predictor variable exhibits a "significant" effect.

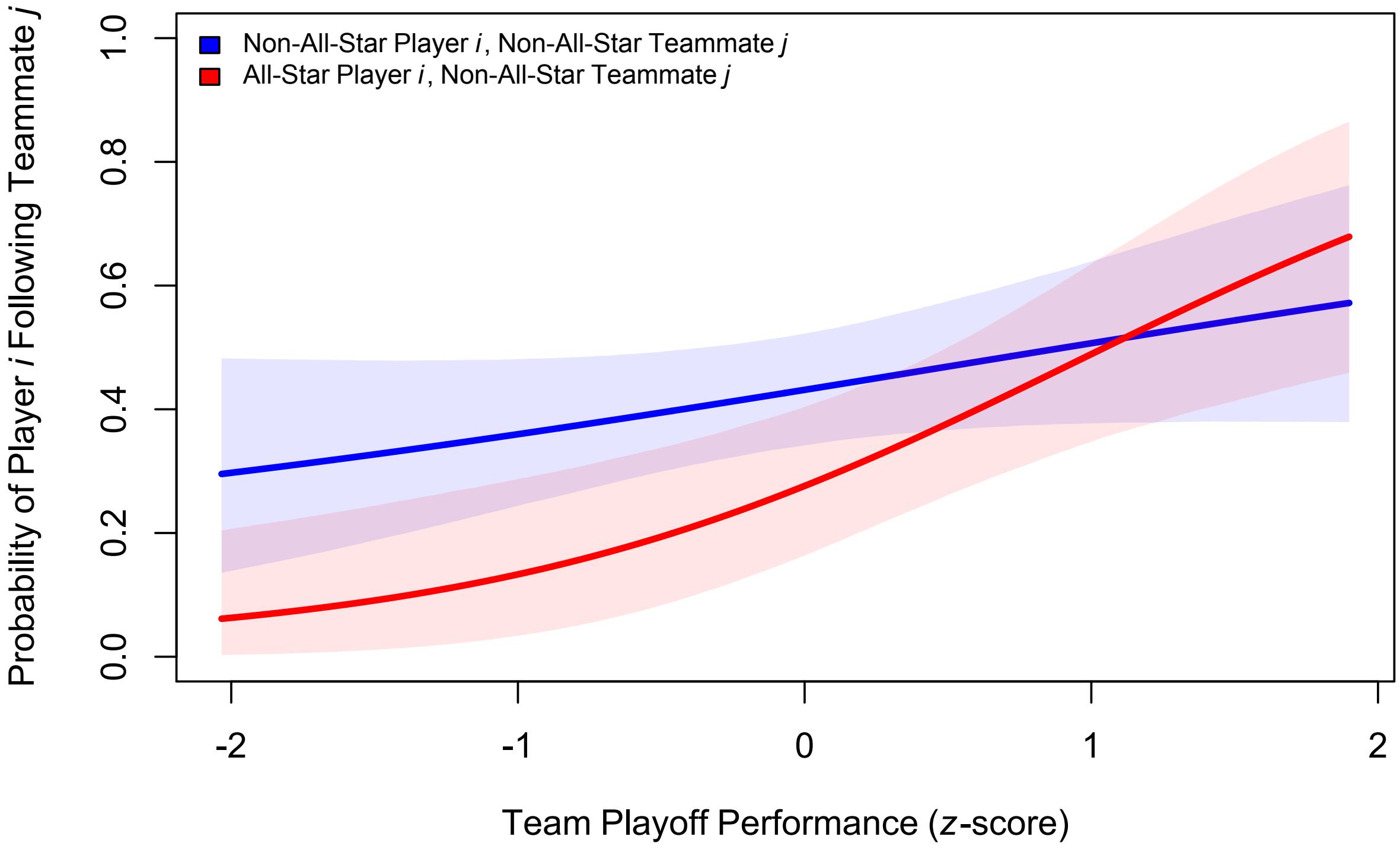

**Figure 4.** Predicted probabilities of Twitter ties as a function of individual *i*'s status as an All-Star and the team's recent performance in the postseason playoffs. In Model 2B, these variables are also interacted with player *j*'s All-Star status, as measured by the number of All-Star games to which the players have been elected. For these predictions, we assume that an All-Star player *i* has been elected to 4 All-Star games whereas a Non-All-Star player has appeared in none. The model predictions assume a league tenure of 4 years for player *i*. Other covariates in the model are held constant at their mean or reference values. The models indicate that compared to All-Star players on successful teams, the All-Star players on poorly performing teams are less likely to follow their teammates on Twitter.

## 3. The Social Relations Model for Count Responses: Single Group

The third version of the SRM demonstrated in this tutorial is appropriate for directed, count data. Like Poisson regression more generally, the model employs a log link function, which constrains predicted values to be positive. That is, the link function precludes models that untenably predict negative values for response variables that cannot be negative, which is typically the case for count data. The model can be written as:

$$y_{i,j} = \text{Poisson}(\lambda_{i,j})$$

$$\ln(\lambda_{i,j}) = \mathbf{x}'_{i,j}\boldsymbol{\beta} + a_i + b_j + e_{i,j}$$

$$\begin{pmatrix} a_i \\ b_i \end{pmatrix} \sim N\left\{\begin{pmatrix} 0 \\ 0 \end{pmatrix}, \begin{pmatrix} \sigma_a^2 & \\ \sigma_{ab} & \sigma_b^2 \end{pmatrix}\right\}, \qquad \rho_{ab} = \frac{\sigma_{ab}}{\sqrt{\sigma_a^2}\sqrt{\sigma_b^2}}$$

$$\begin{pmatrix} e_{i,j} \\ e_{j,i} \end{pmatrix} \sim N\left\{\begin{pmatrix} 0 \\ 0 \end{pmatrix}, \begin{pmatrix} \sigma_e^2 & \\ \sigma_{ee} & \sigma_e^2 \end{pmatrix}\right\}, \qquad \rho_{ee} = \frac{\sigma_{ee}}{\sigma_e^2}$$

$$p_a = \frac{\sigma_a^2}{\sigma_a^2 + \sigma_b^2 + \sigma_e^2}$$

$$p_b = \frac{\sigma_b^2}{\sigma_a^2 + \sigma_b^2 + \sigma_e^2}$$

$$p_e = \frac{\sigma_e^2}{\sigma_a^2 + \sigma_b^2 + \sigma_e^2}$$

The notation for this SRM is largely comparable to the probit model in the first case study, but note that the variance of the random effects, $\sigma_e^2$, is now estimated from the data rather than being constrained to a value of 1.

### 3.1. Case Study 3: Inter-household Food Sharing in a Nicaraguan Community

The third case study replicates an analysis of Koster and Leckie (2014), focusing on food sharing between households in a small community of indigenous Nicaraguans. A total of 25 households were studied for a calendar year, and food diaries were used to track gifts of food (mostly fish or hunted game) from one household to another. The response variable is the number of times that household $i$ gave to household $j$.

Predictor variables include undirected dyad-level variables, such as the distance between households, an "association index" that records how much time members of the households spend together, and inter-household kinship ties, which are modeled with categorical variables. Node-level predictors include the household's productivity in terms of hunting and fishing, a measure of their economic wealth, and the average number of pigs owned by the households during the study period.[12] An additional variable denotes the two households in which the church's pastors reside, as there is a local norm that the pastors should be supported on a limited basis by other members of the community. The model also includes an offset term that captures heterogeneity in the amount of time that both households were simultaneously present in the community. If there is an offset, it needs to be log-transformed prior to its inclusion in the models.

To fit an empty model to these data, follow the steps above to upload and choose the **3_count_one_group_food_sharing.dta** file and the **3_count_one_group.py** template. To fit the empty model (Model 3A), the specification should look like this.

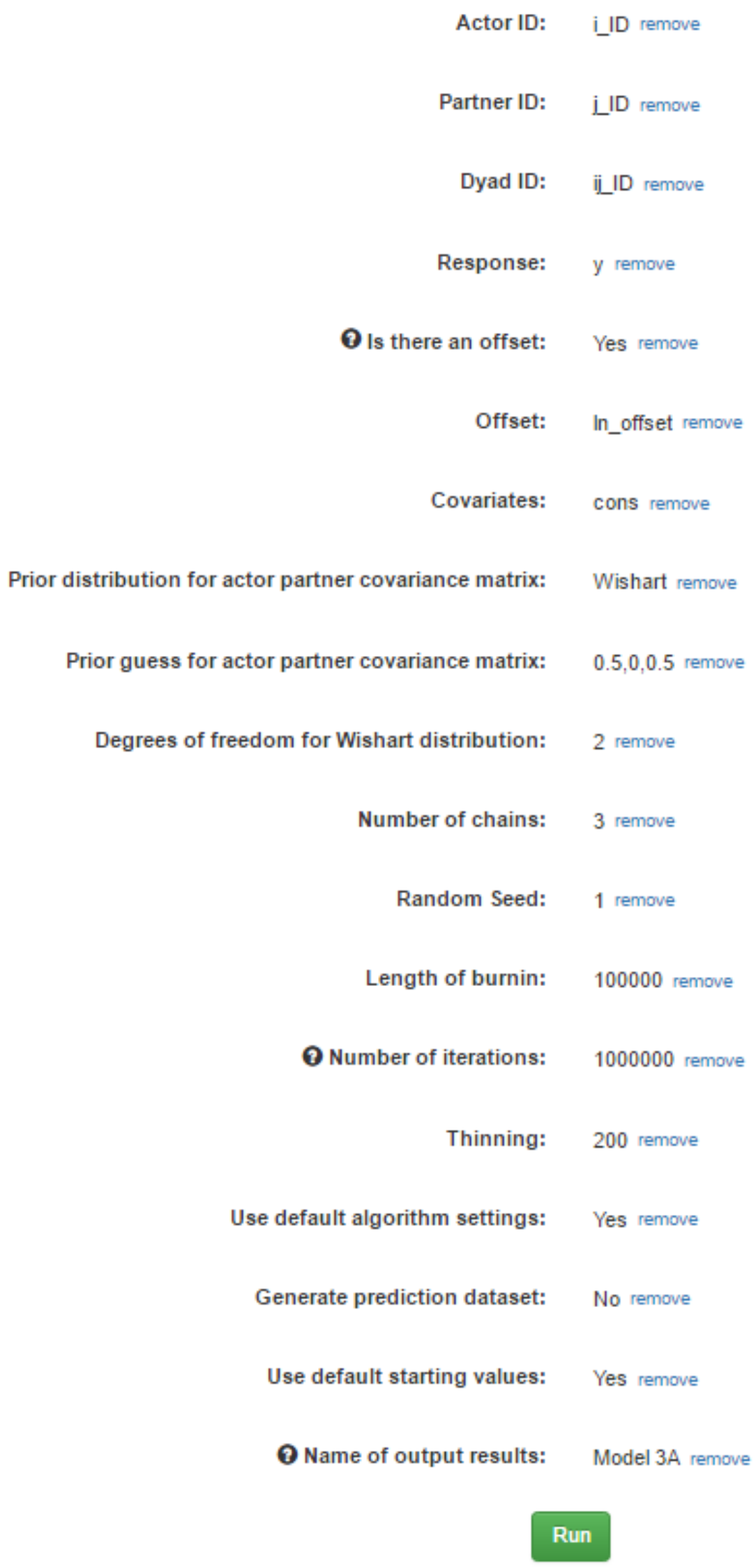

[12] Two of the variables in this dataset exhibit substantial positive skew, the association index and the measure of household hunting productivity. A case could be made for log transforming these variables, but for consistency with the original analysis they are left untransformed. The substantive results change little in models where the variables are transformed.

Two models are presented in Table 5, beginning with an empty model (Model 3A) that includes only the random effects and an intercept, not other predictors. The variance components from this model suggest that over half of the variance lies in the dyadic effects (VPC = 0.58). In addition, households vary more in the role of actors than in the role of partners.[13]

The reciprocity correlations in Table 5 indicate high dyadic reciprocity, which is consistent with game theoretic models of cooperation that emphasize Tit-for-Tat reciprocity and related variants. Interestingly, the generalized reciprocity correlation in Model 3A is negative (−0.49) with a relatively small posterior standard deviation (0.19), suggesting strong statistical support for this negative correlation. The interpretation of this negative correlation is that households that give more food on average tend to receive less from others on average. With the inclusion of covariates in Model 3B, however, this correlation is substantially weaker (0.08) and difficult to distinguish from zero (sd = 0.24).

The explanation for the change in the generalized reciprocity correlation relates to the heterogeneous hunting productivity of the households. That is, the transferred food largely consists of portions of hunted game, and households vary considerably in the amount of game that they harvest. The results suggest a redistributive system of food sharing, with meat flowing from the "haves" to the "have-nots." This is not to suggest, however, that successful hunters give away meat freely to all takers. Rather, meat is given by hunters to preferred recipients, which is evidenced by the importance of the dyadic effects. Nevertheless, the overall trend is toward a redistributive model of food exchange.

Statistically, this example is illustrative because it demonstrates how the estimated variances and correlations in a multilevel model can change substantially with the inclusion of covariates. Among users of multilevel models, a common misconception is that random effects are largely static across models. In the case of the SRM, for instance, users might assume that nodes with high actor-level random intercepts will have similarly high intercepts in all subsequent models. However, random intercepts are not estimated strictly in relation to the intercepts, but rather to the fixed effects portion of the model more generally. Random intercepts that were strongly positive in one model may decrease in magnitude or even flip signs in other models, and vice versa. By extension, the correlations that are based on these random effects will also change, sometimes dramatically.

In this case, the high variance of nodes as actors relates in large part to the variation in hunting productivity. In the empty model (Model 3A), productive households generally have high random intercepts because they have a lot of meat to distribute in the first place. Once the model includes the fixed effect covariate for hunting productivity, the random intercepts no longer reflect this heterogeneous production and the generalized reciprocity correlation diminishes accordingly.[14]

Other covariates in the model tend to conform to predictions. Closer kinship ties predict greater amount of food sharing, and neighbors also show increased propensities to share (hence the negative effect for **Distance**). Wealth has little effect on sharing, but poorer households tend to receive more gifts of food from others. Readers are referred to Koster and Leckie (2014) for further discussion of these models.

---

[13] Note that the VPCs in the Poisson SRM account only for the variance among the higher-level random effects, not the level-1 variation, which is a function of the covariates.

[14] In a similar way, the residuals ($e_{i,j}$ and $e_{j,i}$) can change with the inclusion of covariates, which alters the dyadic reciprocity correlation.

**Table 5.** Parameter estimates from Poisson SRM models fit to count data on food transfers. The values are the means and standard deviations (in parentheses) from the posterior samples.

| | **Variable name** | **Parameter** | **Model 3A** | **Model 3B** |
|---|---|---|---|---|
| $\beta_0$ | Cons | Intercept | 0.68 (0.17) | −0.90 (0.62) |
| $\beta_1$ | ij_mother_offspring | Relatedness type 1 | | 1.51 (0.25) |
| $\beta_2$ | ij_full_sibling_tie | Relatedness type 2 | | 0.96 (0.17) |
| $\beta_3$ | ij_r_25_to_50 | Relatedness type 3 | | 0.30 (0.14) |
| $\beta_4$ | ij_r_10_to_25 | Relatedness type 4 | | 0.02 (0.14) |
| $\beta_5$ | ij_association_index | Association index | | 3.97 (0.64) |
| $\beta_6$ | ij_distance | Distance (log) | | −0.62 (0.08) |
| $\beta_7$ | ij_outlier | Outlier dyad | | 2.47 (0.63) |
| $\beta_8$ | i_hunting | Game – actor | | 0.45 (0.08) |
| $\beta_9$ | j_hunting | Game – partner | | −0.09 (0.07) |
| $\beta_{10}$ | i_fishing | Fishing – actor | | 0.21 (0.72) |
| $\beta_{11}$ | j_fishing | Fishing – partner | | −0.75 (0.59) |
| $\beta_{12}$ | i_pigs | Pigs – actor | | 0.12 (0.05) |
| $\beta_{13}$ | j_pigs | Pigs – partner | | −0.06 (0.04) |
| $\beta_{14}$ | i_wealth | Wealth – actor | | 0.00 (0.02) |
| $\beta_{15}$ | j_wealth | Wealth – partner | | −0.05 (0.02) |
| $\beta_{16}$ | j_pastors | Pastors – partner | | 0.77 (0.36) |
| $\sigma_a^2$ | | Actor variance | 0.69 (0.23) | 0.29 (0.11) |
| $\sigma_b^2$ | | Partner variance | 0.26 (0.10) | 0.17 (0.07) |
| $\sigma_e^2$ | | Dyadic variance | 1.26 (0.14) | 0.30 (0.04) |
| $\rho_{ab}$ | | Generalized reciprocity | −0.49 (0.19) | 0.08 (0.24) |
| $\rho_{ee}$ | | Dyadic reciprocity | 0.95 (0.03) | 0.77 (0.10) |
| | | Actor VPC | 0.31 (0.07) | 0.38 (0.09) |
| | | Partner VPC | 0.12 (0.04) | 0.22 (0.07) |
| | | Dyad VPC | 0.58 (0.07) | 0.40 (0.07) |

Note: Models were using a burn-in of 100,000 iterations and 1,000,000 monitoring iterations.

## 4. The Social Relations Model for Count Responses: Multiple Groups

The fourth version of the SRM in this tutorial is appropriate for directed count data from multiple groups. This is an extension of the Poisson model for the third case study in which an additional random effect, $m_k$, is included for group-level variance. The model is formally notated as follows:

$$y_{ik,jk} = \text{Poisson}(\lambda_{ik,jk})$$

$$\ln(\lambda_{ik,jk}) = \mathbf{x}'_{ik,jk}\boldsymbol{\beta} + m_k + a_{ik} + b_{jk} + e_{ik,jk}$$

$$m_k \sim N(0, \sigma_m^2)$$

$$\begin{pmatrix} a_{ik} \\ b_{ik} \end{pmatrix} \sim N\left\{\begin{pmatrix} 0 \\ 0 \end{pmatrix}, \begin{pmatrix} \sigma_a^2 & \\ \sigma_{ab} & \sigma_b^2 \end{pmatrix}\right\}, \quad \rho_{ab} = \frac{\sigma_{ab}}{\sqrt{\sigma_a^2}\sqrt{\sigma_b^2}}$$

$$\begin{pmatrix} e_{ik,jk} \\ e_{jk,ik} \end{pmatrix} \sim N\left\{\begin{pmatrix} 0 \\ 0 \end{pmatrix}, \begin{pmatrix} \sigma_e^2 & \\ \sigma_{ee} & \sigma_e^2 \end{pmatrix}\right\}, \quad \rho_{ee} = \frac{\sigma_{ee}}{\sigma_e^2}$$

$$p_m = \frac{\sigma_m^2}{\sigma_m^2 + \sigma_a^2 + \sigma_b^2 + \sigma_e^2}$$

$$p_a = \frac{\sigma_a^2}{\sigma_m^2 + \sigma_a^2 + \sigma_b^2 + \sigma_e^2}$$

$$p_b = \frac{\sigma_b^2}{\sigma_m^2 + \sigma_a^2 + \sigma_b^2 + \sigma_e^2}$$

$$p_e = \frac{\sigma_e^2}{\sigma_m^2 + \sigma_a^2 + \sigma_b^2 + \sigma_e^2}$$

As in the second case study, the group-level random effects are assumed to be normally distributed around a mean of zero with a variance estimated from the data ($\sigma_m^2$).

### 4.1. Case Study 4: Co-Sponsorship of Bills in the 108th House of Representatives

The data for the fourth case study stem from research conducted by Fowler (2006), who assembled a large dataset on the co-sponsorship of congressional bills. The present subset of the data examines only the co-sponsorship of bills in the House of Representatives in the 108th Congress, which held sessions from 2003 to 2005. Fowler (2006) provides a brief explanation of co-sponsorship, noting that when a legislative bill is advanced by a representative, other representatives can express support for the bill by signing it as a co-sponsor. This support has value, and legislators sometimes expend considerable effort seeking co-sponsors for their bills.

In turn, co-sponsoring individuals might expect that the colleagues they support could in turn co-sponsor their own bills in the future, thus exhibiting positive dyadic reciprocity.

Although Fowler (2006) presents the full data on co-sponsoring networks in the 108th House of Representatives, this tutorial focuses on a subset of the data, specifically on the co-sponsorship of bills among legislators who represent the same state. Some states, however, have only one representative, and these individuals were excluded from the dataset because they were not members of any within-state dyads. One legislator sponsored no bills and was also removed from the dataset. The dataset examined here thus includes a total of 7,748 directed relationships by 431 congresspersons distributed among 44 states. The dataset is called **4_Count_multi_group_congress.dta** and models can be fit using the **4_Count_multi_group.py** template. As a clarifying note, the actors represent potential senders of co-sponsoring ties to partners, who are the representatives who sponsored bills. In other words, the response variable is the number of ties that *i* has signed on to co-sponsor legislation by *j*. The number of bills initially sponsored by *j* shows considerable heterogeneity, meaning that potential co-sponsors have varying opportunities to co-sponsor the bills of their peers. Therefore, the dataset also includes an offset variable, corresponding to the number of bills sponsored by individual *j* during the congressional sessions. Instead of modeling the raw counts of co-sponsorship, the offset changes the interpretation of the model to signify the rate of co-sponsorship per bill.[15]

The political party, either Democrat or Republican, of each representative is known.[16] We anticipated that representatives would be more likely to co-sponsor bills from members of their own party, so binary variables were created to denote whether *i* and *j* are Democrats, respectively. As in previous case studies, we then created an interaction term by multiplying these variables, where a 1 now indicates that both *i* and *j* are Democrats. We also conjectured that there would be greater co-sponsorship support in low-population states, which have fewer representatives and plausibly greater convergence in legislative interests. A state-level variable, **reps_perc_max**, was therefore created to denote the number of representatives in each state relative to the maximum value of 53 representatives (in California). Note that this proportional rescaling of the variable to have a maximum of 1 was done in large part to improve estimation – parameters with large values often do not mix well. In general, researchers using the SRM templates should consider rescaling variables (such as *z*-score transformations) as a remedy for poorly-mixing chains.

After specifying an empty model (Model 4A), we include the main effects for **reps_perc_max**, the political party of *i* and *j*, and the interaction term in Model 4B. Finally, we consider the possibility that partisan co-sponsorship could be moderated by the size of the state.

---

[15] This dataset was selected for the case study because distribution of the response variable in this analysis exhibits the positive skew that characterizes count data. Nevertheless, Poisson models for count data tend to be ideal when the maximum number of possible events is unknown. In the case of congressional co-sponsoring, however, the possibility to co-sponsor is conditional on the advancement of a bill by a legislator. The number of bills sponsored by each representative is known, so the data could alternately be recast as proportions that would lend themselves to a binomial regression version of the SRM. To our knowledge, this version of the SRM has not been developed. The merit of an offset to represent the variation in sponsored bills by representative *j* is debatable, though it is worth noting that binomial and Poisson parameterizations typically yield comparable inferences (McElreath, 2015:326-327).

[16] One representative, Bernie Sanders, was a political independent. However, because he was the only representative from Vermont, Sanders does not appear as a node in this dataset.

We therefore create terms for a three-way interaction between all of the main effects. This latter model (Model 4C) appears as follows when fitted in Stat-JR:

| | |
|---|---|
| Group ID | k_ID remove |
| Actor ID: | i_ID remove |
| Partner ID: | j_ID remove |
| Dyad ID: | ij_ID remove |
| Response: | y remove |
| Is there an offset: | Yes remove |
| Offset: | log_offset remove |
| Covariates: | cons,reps_perc_max,i_democrat,j_democrat,int_ij_democrat,int_reps_i_democrat,int_reps_j_democrat,int_reps_ij_democrat remove |
| Prior distribution for actor partner covariance matrix: | Wishart remove |
| Prior guess for actor partner covariance matrix: | 0.5,0,0.5 remove |
| Degrees of freedom for Wishart distribution: | 2 remove |
| Number of chains: | 3 remove |
| Random Seed: | 1 remove |
| Length of burnin: | 100000 remove |
| Number of iterations: | 2000000 remove |
| Thinning: | 400 remove |
| Use default algorithm settings: | Yes remove |
| Generate prediction dataset: | No remove |
| Use default starting values: | Yes remove |
| Name of output results: | Model 4C remove |
| | Run |

In general, despite fitting three chains of 2 million iterations, the chains for some parameters exhibit poor mixing. That is especially true of the chains relating to the dyadic variance and reciprocity, which is attributable in part to the very high reciprocity. The estimate for dyadic reciprocity consistently approaches the boundary in the empty model ($\rho_{ee} = 0.99$), which leaves little parameter space for the algorithm to explore. It would be wise to consider running longer chains to obtain more robust estimates, a caveat to bear in mind for the interpretation of the models presented here.

Table 6 presents the results of the three models. As noted, the dyadic reciprocity is high across all of the models, but there is little evidence for generalized reciprocity in co-sponsorship. Interestingly, the variance partition coefficients are dominated by partner-level variance, and that is especially true in Model 4B and Model 4C, presumably because party affiliations explain much of the dyadic variance and **Representation** explains group-level variance. The interpretation is that some partners successfully attract many co-sponsorships from their peers for reasons that are not captured by variables in the model.

**Table 6.** Parameter estimates from Poisson SRM models fit to count data on congressional co-sponsoring of bills from the 108th House of Representatives. In this analysis, the actor is a representative who potentially co-sponsors bills that were initiated by the partner. The values are the means and standard deviations (in parentheses) from the posterior samples for each parameter.

| **Variable name** | **Description** | **Model 4A** | **Model 4B** | **Model 4C** |
|---|---|---|---|---|
| cons | Intercept | -2.05 | -1.63 | -1.53 |
| | | (0.08) | (0.08) | (0.09) |
| reps_perc_max | Representation | | -1.09 | -1.38 |
| | | | (0.17) | (0.22) |
| i_democrat | Democrat actor | | -0.41 | -0.39 |
| | | | (0.04) | (0.07) |
| j_democrat | Democrat partner | | -0.74 | -0.55 |
| | | | (0.08) | (0.13) |
| int_ij_democrat | Interaction: Actor Democrat * | | 1.56 | 0.97 |
| | Partner Democrat | | (0.04) | (0.08) |
| int_reps_i_democrat | Interaction: Representation * | | | 0.02 |
| | Democrat actor | | | (0.13) |
| int_reps_j_democrat | Interaction: Representation | | | -0.24 |
| | Democrat partner | | | (0.27) |
| int_reps_ij_democrat | Interaction: Representation * Actor | | | 1.06 |
| | Democrat * Partner Democrat | | | (0.14) |
| | Group-level variance | 0.10 | 0.01 | 0.01 |
| | | (0.05) | (0.01) | (0.01) |
| | Actor variance | 0.10 | 0.06 | 0.06 |
| | | (0.01) | (0.01) | (0.01) |
| | Partner variance | 0.54 | 0.53 | 0.53 |
| | | (0.05) | (0.04) | (0.04) |
| | Dyadic variance | 0.12 | 0.02 | 0.02 |
| | | (0.01) | (0.01) | (0.01) |
| | Generalized reciprocity | 0.07 | 0.02 | 0.00 |
| | | (0.07) | (0.07) | (0.07) |
| | Dyadic reciprocity | 0.99 | 0.91 | 0.91 |
| | | (0.01) | (0.09) | (0.11) |
| | Group-level VPC | 0.11 | 0.01 | 0.01 |
| | | (0.05) | (0.02) | (0.02) |
| | Actor VPC | 0.12 | 0.10 | 0.09 |
| | | (0.01) | (0.01) | (0.01) |
| | Partner VPC | 0.63 | 0.85 | 0.86 |
| | | (0.04) | (0.02) | (0.02) |
| | Dyad VPC | 0.14 | 0.04 | 0.03 |
| | | (0.01) | (0.01) | (0.01) |

Note: Models were using a burn-in of 100,000 iterations and 2,000,000 monitoring iterations.

As predicted, there is evidence for party allegiance, as both Democrats and Republicans co-sponsor their peers' bills at greater rates than across party lines. Also, as predicted, the population size of the state (as proxied by **reps_perc_max**) exhibits a negative effect on the co-sponsoring rate, as seen in Model 4B ($\beta = 0.82$). Note that the coefficients are on the log scale, so the exponential function is used to calculate predictions on the scale of the response.

The last model, Model 4C, exhibits a significantly positive three-way interaction effect ($\beta = 1.06$). To interpret this effect, we use the posterior mean coefficients to plot the cluster-specific predictions as a function of state population for two combinations of actors and partners: one set of predictions assumes that both the actor and partner are Republicans, and the other set of predictions assumes that both individuals are Democrats.[17] As seen in Figure 5, the model suggests that party affiliation moderates the effect of population size on co-sponsorship rates. In small states, both Republicans and Democrats show equal penchants for co-sponsoring legislation that was initiated by their in-state political peers. As population size increases, however, the slope for Republicans is significantly lower than the slope for Democrats, suggesting that Democrats in larger states extend more support to other members of their party than Republicans do.

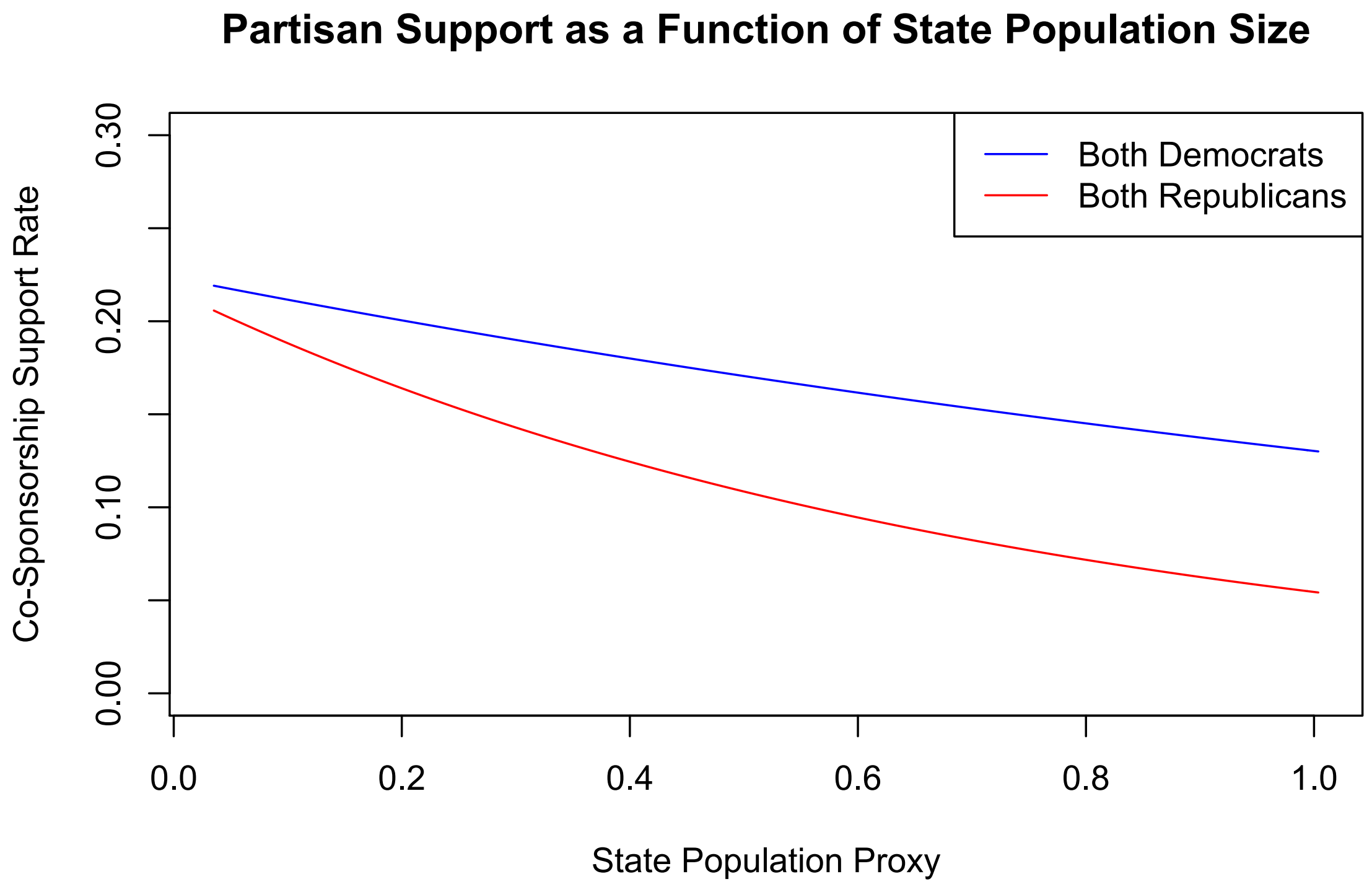

**Figure 5.** Cluster-specific model predictions from Model 4C, showing two combinations of actor-partner dyads. One combination assumes that both the actor and partner are Democrats, and the other assumes that both are Republicans. Predictions for other cross-party combinations of actors and partners are not displayed.

[17] It would be possible to simulate predictions for each sample in the posterior, as in Case Study 2. For instance, the script in Appendix B could be modified to calculate predictions using the exponential function.

## 5. The Social Relations Model for Continuous Responses: Single Group

The fifth version of the SRM in this tutorial is appropriate for directed, continuous data from a single group. The model is formally notated as follows:

$$y_{i,j} = \mathbf{x}'_{i,j}\boldsymbol{\beta} + a_i + b_j + e_{i,j}$$

$$\begin{pmatrix} a_i \\ b_i \end{pmatrix} \sim N\left\{\begin{pmatrix} 0 \\ 0 \end{pmatrix}, \begin{pmatrix} \sigma_a^2 & \\ \sigma_{ab} & \sigma_b^2 \end{pmatrix}\right\}, \qquad \rho_{ab} = \frac{\sigma_{ab}}{\sqrt{\sigma_a^2}\sqrt{\sigma_b^2}}$$

$$\begin{pmatrix} e_{i,j} \\ e_{j,i} \end{pmatrix} \sim N\left\{\begin{pmatrix} 0 \\ 0 \end{pmatrix}, \begin{pmatrix} \sigma_e^2 & \\ \sigma_{ee} & \sigma_e^2 \end{pmatrix}\right\}, \qquad \rho_{ee} = \frac{\sigma_{ee}}{\sqrt{\sigma_e^2}\sqrt{\sigma_e^2}}$$

$$p_a = \frac{\sigma_a^2}{\sigma_a^2 + \sigma_b^2 + \sigma_e^2}$$

$$p_b = \frac{\sigma_b^2}{\sigma_a^2 + \sigma_b^2 + \sigma_e^2}$$

$$p_e = \frac{\sigma_e^2}{\sigma_a^2 + \sigma_b^2 + \sigma_e^2}$$

The models and notation here are distinguished from previous models by the absence of a link function.

### 5.1. Case Study 5: International Trade Data

The data for the fifth case study stem from a dataset on international trade relations among the 30 countries with the highest gross domestic product (GDP), as described by Ward and Hoff (2007). The data are drawn from a built-in dataset in Peter Hoff's *amen* package, and the presentation largely follows the treatment in the tutorial for that package (Hoff, 2015).[18] The response variable is the log-transformed value of exports from country *i* to country *j*. Predictors include two dyadic covariates, namely the geographic distance between the countries (**distance**) and the number of inter-governmental organizations to which both countries belong (**shared_igos**). The dataset also includes several node-level predictors, including the population size of the respective countries on the logarithmic scale (**i_population_log** and

---

[18] The *amen* package has a number of model extensions that might be of interest to network researchers. In addition to estimating the basic SRM, the package allows for random effects that reflect triadic dependencies and other latent structures in the network. This package also addresses the problem of censoring that occasionally characterizes organizational research on social networks. That is, researchers using a "name generator" method sometimes invite participants to list alters up to a maximum number. Once an informant reaches the maximum, then any remaining unnamed alters are censored because it is unclear if ego would have nominated that individual in the absence of the constraint. The *amen* package allows for the estimation of models that account for the uncertainty that stems from this censoring.

**j_population_log**), their gross domestic product on the logarithmic scale (**i_gdp_log** and **j_gdp_log**), a measure of their commitment to democratic ideals (**i_polity_10** and **j_polity_10**). As in several of our previous case studies, we use these latter two node-level variables to generate an interaction term.

The dataset (**5_Continuous_one_group_trade.dta**) is modeled with the fifth SRM template (**5_Continuous_one_group.py**). The empty model (Model 5A) appears as follows when specified in Stat-JR:

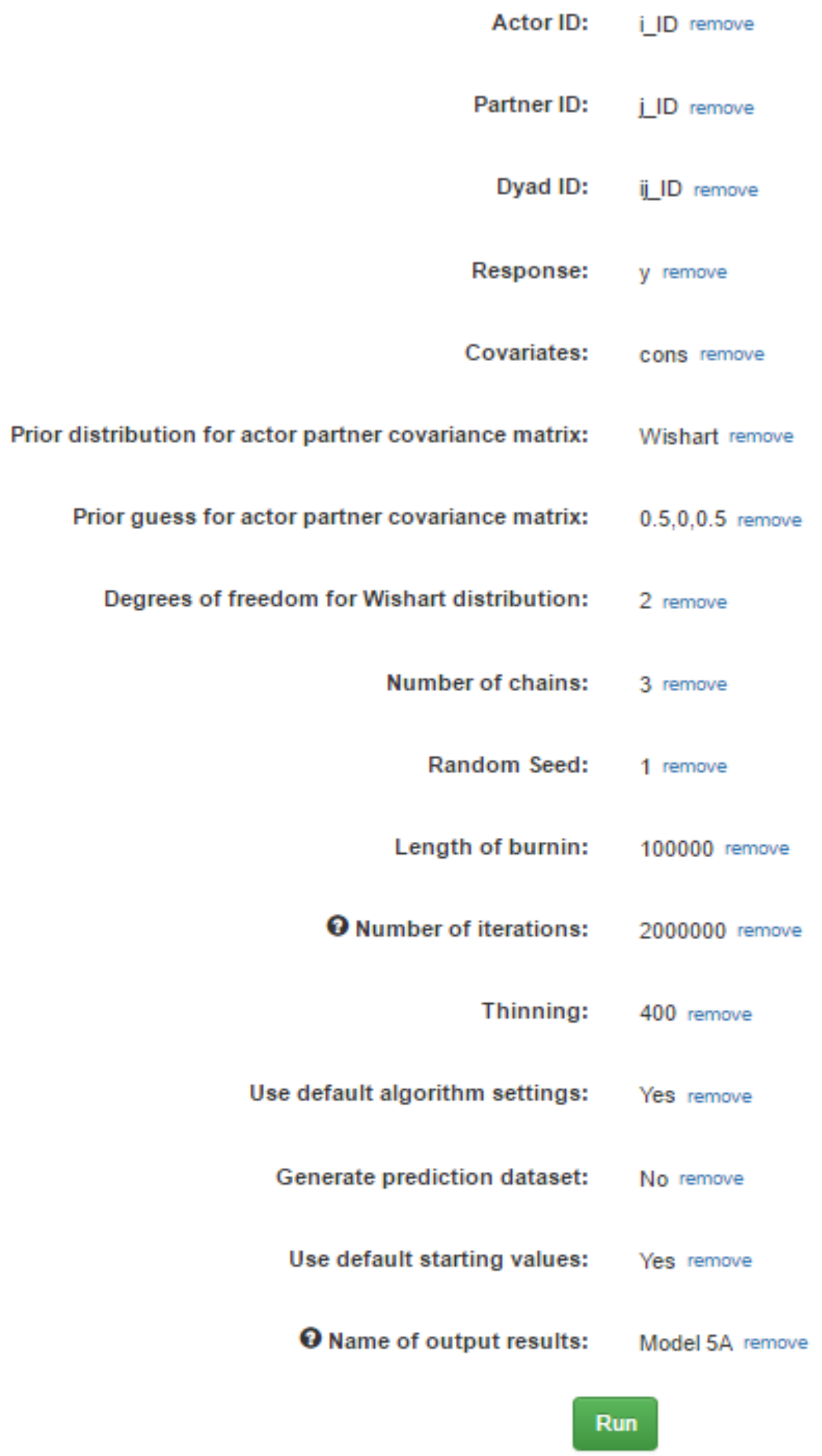

Table 7 includes the parameter estimates from the three models that were fitted to the data. The VPCs suggest that all three variance components contribute roughly equally to variation in the response variable (values from 0.30 to 0.36). Both generalized reciprocity (0.85) and dyadic reciprocity (0.86) are high, though they diminish moderately in subsequent models (see Case Study 3).

**Table 7.** Parameter estimates from SRM models fit to data on the value of exports between countries. The values are the means and standard deviations (in parentheses) from the posterior samples. The response variable is the log-transformed value of exports from country *i* to country *j*.

| Variable name | Parameter | Model 5A | Model 5B | Model 5C |
|---|---|---|---|---|
| cons | Intercept | 0.68 | -4.41 | -4.49 |
| | | (0.19) | (0.89) | (0.92) |
| distance | Distance | | -0.03 | -0.03 |
| | | | (0.01) | (0.01) |
| shared_igos | Shared inter-government | | 0.02 | 0.03 |
| | organizations | | (0.00) | (0.00) |
| i_population_log | Actor population size | | -0.28 | -0.27 |
| | | | (0.10) | (0.10) |
| j_population_log | Partner population size | | -0.24 | -0.23 |
| | | | (0.09) | (0.1) |
| i_gdp_log | Actor gross domestic product | | 0.55 | 0.53 |
| | | | (0.11) | (0.12) |
| j_gdp_log | Partner gross domestic product | | 0.51 | 0.50 |
| | | | (0.11) | (0.11) |
| i_polity_10 | Actor democracy score | | -0.23 | -0.12 |
| | | | (0.15) | (0.16) |
| j_polity_10 | Partner democracy score | | -0.13 | -0.02 |
| | | | (0.14) | (0.15) |
| ij_polity_10_int | Interaction: Actor democracy * | | | -0.19 |
| | Partner democracy | | | (0.06) |
| | Actor variance | 0.29 | 0.15 | 0.17 |
| | | (0.08) | (0.05) | (0.05) |
| | Partner variance | 0.28 | 0.14 | 0.15 |
| | | (0.08) | (0.05) | (0.05) |
| | Dyadic variance | 0.24 | 0.14 | 0.14 |
| | | (0.02) | (0.01) | (0.01) |
| | Generalized reciprocity | 0.85 | 0.68 | 0.71 |
| | | (0.05) | (0.11) | (0.1) |
| | Dyadic reciprocity | 0.86 | 0.77 | 0.76 |
| | | (0.01) | (0.02) | (0.02) |
| | Actor VPC | 0.36 | 0.35 | 0.36 |
| | | (0.04) | (0.06) | (0.06) |
| | Partner VPC | 0.34 | 0.31 | 0.33 |
| | | (0.04) | (0.05) | (0.05) |
| | Dyad VPC | 0.30 | 0.34 | 0.32 |
| | | (0.05) | (0.06) | (0.06) |

Note: Models were using a burn-in of 100,000 iterations and 2,000,000 monitoring iterations.

In terms of covariates, Model 5B shows that proximate countries exchange more ($\beta = -0.03$), and countries that partner in international organizations also exchange more ($\beta = 0.02$). Controlling for other parameters, larger countries both export ($\beta = -0.28$). and receive less ($\beta = -0.24$). Intuitively, countries with large GDP export more than their peers ($\beta = 0.55$), and they also import more ($\beta = 0.51$). The democratic orientation of the countries seems to have little effect on trade relations for actors ($\beta = -0.23$) or partners ($\beta = -0.13$).

In Model 5C, however, the interaction of the respective nations' democracy scores exhibits a negative effect ($\beta = -0.19$). This result implies that as both countries exhibit more democratic ideals, the volume of trade between them declines. It is worth noting, however, that the interaction term is moderately correlated with the **shared_igos** variable, and when this latter variable is omitted from the model, then the interaction term no longer exhibits a noteworthy effect. The data presented here also do not include variables that potentially explain variation in trade relations, such as the recent history of conflicts between the countries (Ward and Hoff, 2007).

## 6. The Social Relations Model for Continuous Responses: Multiple Groups

The sixth version of the SRM in this tutorial is appropriate for directed, continuous data from multiple groups, a research design that is commonly used by social psychologists. The model adds a group-level random effect, $m_k$, to account for variation in network density across groups:

$$y_{ik,jk} = \mathbf{x}'_{ik,jk}\boldsymbol{\beta} + m_k + a_{ik} + b_{jk} + e_{ik,jk}$$

$$m_k \sim N(0, \sigma_m^2)$$

$$\begin{pmatrix} a_{ik} \\ b_{ik} \end{pmatrix} \sim N\left\{\begin{pmatrix} 0 \\ 0 \end{pmatrix}, \begin{pmatrix} \sigma_a^2 & \\ \sigma_{ab} & \sigma_b^2 \end{pmatrix}\right\}, \qquad \rho_{ab} = \frac{\sigma_{ab}}{\sqrt{\sigma_a^2}\sqrt{\sigma_b^2}}$$

$$\begin{pmatrix} e_{ik,jk} \\ e_{jk,ik} \end{pmatrix} \sim N\left\{\begin{pmatrix} 0 \\ 0 \end{pmatrix}, \begin{pmatrix} \sigma_e^2 & \\ \sigma_{ee} & \sigma_e^2 \end{pmatrix}\right\}, \qquad \rho_{ee} = \frac{\sigma_{ee}}{\sqrt{\sigma_e^2}\sqrt{\sigma_e^2}}$$

$$p_m = \frac{\sigma_m^2}{\sigma_m^2 + \sigma_a^2 + \sigma_b^2 + \sigma_e^2}$$

$$p_a = \frac{\sigma_a^2}{\sigma_m^2 + \sigma_a^2 + \sigma_b^2 + \sigma_e^2}$$

$$p_b = \frac{\sigma_b^2}{\sigma_m^2 + \sigma_a^2 + \sigma_b^2 + \sigma_e^2}$$

$$p_e = \frac{\sigma_e^2}{\sigma_m^2 + \sigma_a^2 + \sigma_b^2 + \sigma_e^2}$$

### 6.1. Case Study 6: Peer Rankings of Leadership in Task Groups

The data for the sixth case study are from a study of leadership conducted by Kennedy et al. (2013). Participants in the study were 140 university students and staff members, who were independently divided into 4-person groups (*n* = 35) to work on judgment and general knowledge tasks. Following this initial phase of the research (Phase 1 of the study), the participants privately rated all members in the group, including themselves, in terms of how much leadership they displayed, on a scale of 1 (weak leadership) to 5 (strong leadership).[19] The response variable is the peer rating of actor *i* of partner *j*. Predictor variables are based on the self-ratings of leadership, with **i_self_rating** representing the actor's self-evaluation and **j_self_rating** representing the partner's self-evaluation of leadership.[20]

---

[19] As an alternative to gaussian models, it might be worthwhile to consider an ordered probit Social Relations Model for scaled data.

[20] The report by Kennedy et al. (2013) provides insight into several additional predictor variables that are not included in this tutorial.

The dataset (**6_Continuous_multi_group_leadership.dta**) is modeled with the sixth SRM template (**6_Continuous_multi_group.py**). The empty model (Model 6A) appears as follows when specified in Stat-JR:

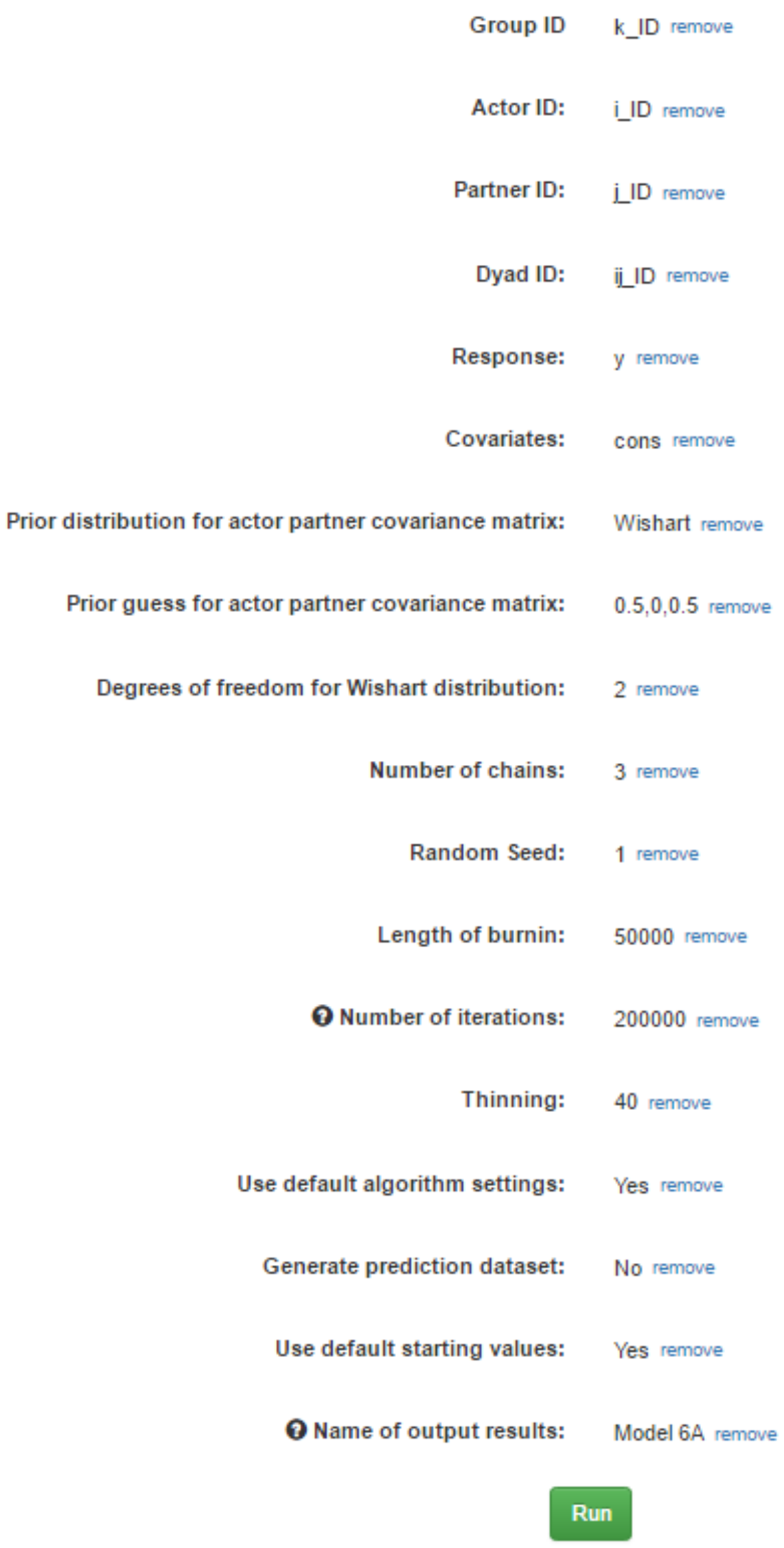

Table 8 includes the parameter estimates from two models that were fitted to the data. In the empty model (Model 6A), actors vary relatively little in the ratings they assign to their peers ($\sigma_a^2 = 0.14$). The variance partition coefficients instead show that partner effects and dyadic effects are the primary source of variation in the dataset. Group-level variance is essentially absent, indicating that groups do not systematically assign higher or lower ratings to their peers. The reciprocity correlations are largely indistinguishable from a zero correlation (although there is a modestly negative generalized reciprocity correlation with the inclusion of covariates in Model 6B).

The self-ratings are both significant predictors of peer evaluations. Actors who rate themselves highly for leadership also rate their peers relatively highly ($\beta = 0.18$). Also, partners who rate themselves highly are rated by their peers as providers of stronger leadership ($\beta = 0.45$).

**Table 8.** Parameter estimates from SRM models fit to data on peer ratings of leadership in small task groups. The values are the means and standard deviations (in parentheses) from the posterior samples. The response variable is the rating by individual *i* of individual *j*.

| Variable name | Parameter | Model 6A | Model 6B |
|---|---|---|---|
| cons | Intercept | 3.65 (0.09) | 1.28 (0.31) |
| i_self_rating | Actor self-rating | | 0.18 (0.05) |
| j_self_rating | Partner self-rating | | 0.45 (0.07) |
| | Group-level variance | 0.01 (0.02) | 0.01 (0.02) |
| | Actor variance | 0.14 (0.04) | 0.13 (0.04) |
| | Partner variance | 0.66 (0.11) | 0.47 (0.09) |
| | Dyadic variance | 0.58 (0.06) | 0.55 (0.05) |
| | Generalized reciprocity | −0.01 (0.17) | −0.26 (0.16) |
| | Dyadic reciprocity | −0.01 (0.10) | −0.06 (0.10) |
| | Group-level VPC | 0.01 (0.01) | 0.01 (0.01) |
| | Actor VPC | 0.10 (0.03) | 0.12 (0.03) |
| | Partner VPC | 0.47 (0.05) | 0.40 (0.05) |
| | Dyad VPC | 0.42 (0.05) | 0.47 (0.05) |

Note: Models were using a burn-in of 50,000 iterations and 200,000 monitoring iterations.

# Appendix A

Key for Interpreting Parameters in SRM Templates

The following keys provide clarification on the model parameters provided by the Stat-JR templates.

**SRM Template**: *1_Binary_one_group.py*

| Reported parameter | Parameter | Notes |
|---|---|---|
| sigma2b1 | Partner variance | |
| pb1 | Partner VPC | |
| rhoe1e1 | Dyadic reciprocity | |
| sigma2a1 | Actor variance | |
| pe1 | Dyad VPC | |
| sigma2r | Sum of variances | This is a placeholder parameter used only for calculating VPCs. |
| pa1 | Actor VPC | |
| deviance | Model deviance | |
| rhoa1b1 | Generalized reciprocity | |

**SRM Template**: *2_Binary_multi_group.py*

| Reported parameter | Parameter | Notes |
|---|---|---|
| pa1 | Actor variance | |
| sigma2b1 | Partner variance | |
| pb1 | Partner VPC | |
| sigma2m | Group-level variance | |
| sigma2a1 | Actor-level variance | |
| rhoe1e1 | Dyadic reciprocity | |
| pe1 | Dyad VPC | |
| sigma2r | Sum of variances | This is a placeholder parameter used only for calculating VPCs. |
| deviance | Model deviance | |
| pm | Group-level VPC | |
| rhoa1b1 | Generalized reciprocity | |

**SRM Template**: *3_Count_one_group.py*

| Reported parameter | Parameter | Notes |
|---|---|---|
| pa1 | Actor VPC | |
| sigma2b1 | Partner variance | |
| pb1 | Partner VPC | |
| sigma2e1 | Dyadic variance | |
| sigma2a1 | Actor variance | |
| pe1 | Dyad VPC | |
| sigma2r | Sum of variances | This is a placeholder parameter used only for calculating VPCs. |
| deviance | Model deviance | |
| rhoe1e1 | Dyadic reciprocity | |
| rhoa1b1 | Generalized reciprocity | |

**SRM Template**: *4_Count_multi_group.py*

| Reported parameter | Parameter | Notes |
|---|---|---|
| sigma2a1 | Actor variance | |
| rhoa1b1 | Generalized reciprocity | |
| pb1 | Partner VPC | |
| pm | Group-level VPC | |
| sigma2b1 | Partner variance | |
| sigma2m | Group-level variance | |
| pa1 | Actor VPC | |
| deviance | Model deviance | |
| pe1 | Dyad VPC | |
| sigma2e1 | Dyadic variance | |
| sigma2r | Sum of variances | This is a placeholder parameter used only for calculating VPCs. |
| rhoe1e1 | Dyadic reciprocity | |

**SRM Template**: *5_Continuous_one_group.py*

| Reported parameter | Parameter | Notes |
|---|---|---|
| sigma2b1 | Partner variance | |
| pb1 | Partner VPC | |
| sigma2e1 | Dyadic variance | |
| sigma2a1 | Actor variance | |
| pe1 | Dyad VPC | |
| sigma2r | Sum of variances | This is a placeholder parameter used only for calculating VPCs. |
| pa1 | Actor VPC | |
| deviance | Model deviance | |
| rhoe1e1 | Dyadic reciprocity | |
| rhoa1b1 | Generalized reciprocity | |

**SRM Template**: *6_Continuous_multi_group.py*

| Reported parameter | Parameter | Notes |
|---|---|---|
| sigma2a1 | Actor variance | |
| rhoa1b1 | Generalized reciprocity | |
| pb1 | Partner VPC | |
| pm | Group-level VPC | |
| sigma2b1 | Partner variance | |
| sigma2m | Group-level variance | |
| pa1 | Actor VPC | |
| deviance | Model deviance | |
| pe1 | Dyad VPC | |
| sigma2e1 | Dyadic variance | |
| sigma2r | Sum of variances | This is a placeholder parameter used only for calculating VPCs. |
| rhoe1e1 | Generalized reciprocity | |

## Appendix B

R Script to Plot Predictions of Model 2B

```
## The following script generates "cluster specific" predictions
## from the posterior samples of the fixed effects from Model 2C.
## The script relies on two R packages, foreign and rethinking.

library(foreign) ## Available on CRAN
library(rethinking) ## Available here: http://xcelab.net/rm/software/

## Reads in raw data, which will be used to generate the scatterplot.
## These same data were used as input for the model in Stat-JR.
## Users will need to point R to the file in their working directory.
d <- read.dta (file= "./2_Binary_multi_group_NBA_twitter.dta")

## Creates a proportional variable, y, for each of the players in the
## dataset, reflecting the percentage of teammates who follow him.
d.j <- aggregate ( y ~ j_ID + j_log_salary, data = d, FUN = mean)

## This command reads in the posterior samples generated by Stat-JR.
## These samples can be downloaded from the Stat-JR interface.
post <- read.dta ( file = "./Model 2C")

## This is the function that generates posterior predictions
## from a data frame called "pred_data_j" containing values
## for which predictions will be generated.
p.pred.j <- function(i) {
 with ( pred_data_j , pnorm(
 post$beta_0 +
     post$beta_1 * mean(d$i_age) +
     post$beta_2 * mean(d$j_age) +
     post$beta_3 * mean(d$i_height) +
     post$beta_4 * mean(d$j_height) +
     post$beta_5 * mean(d$i_time) +
     post$beta_6 * mean(d$j_time) +
     post$beta_7 * mean(d$i_log_tenure) +
     post$beta_8 * mean(d$j_log_tenure) +
     post$beta_9 * i_log_salary[i] +
     post$beta_10 * j_log_salary[i] +
     post$beta_11 * 0 +
     post$beta_12 * mean(d$k_sqrt_wins) +
     post$beta_13 * ij_salary_interaction[i]
      ) ) }

## Creation of a sequence of alter salary values to predict for a low-
```

```
 salary ego.
j_log_salary <- seq (from = min(d$j_log_salary), to =
 max(d$j_log_salary), by = .005)

## Convert to a data frame.
pred_data_j <- as.data.frame (j_log_salary)

## Quantile function returns salary of players at 10% percentile of
 salary distribution
pred_data_j$i_log_salary <- rep (as.numeric(quantile(d.j$j_log_salary,
 .1)), length.out = nrow(pred_data_j))

## Adding the interaction term to the prediction data frame.
pred_data_j$ij_salary_interaction <- pred_data_j$i_log_salary *
 pred_data_j$j_log_salary

## Generating predictions by calling the function
p.low <- sapply ( 1:nrow(pred_data_j) , p.pred.j)
p.mean.low.salary <- apply ( p.low, 2 , mean )
percentiles.low.salary <- apply ( p.low , 2, quantile, c(.025, .975))

## Plotting predictions
plot ( d.j$j_log_salary, d.j$y, main = "Interaction of Actor and
 Partner Salaries", ylab = "Probability of Actor Following Partner",
 xlab = "Partner's Salary (Millions of Dollars)", xaxt = "n")

legend ( min(d$j_log_salary), 1, c ("High Salary Actor", "Low Salary
 Actor"), lty = c(1,4), cex = 0.8)
salary.labels <- c ( 0.1, 1, 5, 20)

axis (1, at = log (c (0.1, 1, 5, 20)), labels = salary.labels)
lines ( pred_data_j$j_log_salary, p.mean.low.salary, lty = 4)
shade (percentiles.low.salary, pred_data_j$j_log_salary)

## Leave the plot device open so that additional predictions can
## be plotted for high-salary egos.

## This is housekeeping to make sure we do not erroneously reuse
## the previous data frame for low-salary egos.
rm (j_log_salary)
rm (pred_data_j)

## Now the process is repeated for high-salary egos.
j_log_salary <- seq (from = min(d$j_log_salary), to =
 max(d$j_log_salary), by = .005)
```

```
pred_data_j <- as.data.frame (j_log_salary)

## 90th percentile of salary distribution
pred_data_j$i_log_salary <- rep (as.numeric(quantile(d.j$j_log_salary,
 0.9)), length.out = nrow(pred_data_j))

pred_data_j$ij_salary_interaction <- pred_data_j$i_log_salary *
 pred_data_j$j_log_salary

p.high <- sapply ( 1:nrow(pred_data_j) , p.pred.j)
p.mean.high <- apply ( p.high, 2 , mean )
percentiles.high.salary <- apply (p.high, 2, quantile, c(.025, .975))

## Assuming the plot device is still open, the additional prediction
## lines and intervals will be added to the figure.
lines ( pred_data_j$j_log_salary, p.mean.high)
shade (percentiles.high.salary, pred_data_j$j_log_salary)

## The figure should resemble Figure 4 in this tutorial.
```